\documentclass[aps,a4paper,preprint, superscriptaddress,12pt,preprintnumbers,floatfix,nofootinbib,amsmath,amssymb]{revtex4}
\usepackage{amstext,amssymb}
\usepackage{amsmath}
\usepackage{graphicx}
\usepackage{dcolumn}
\usepackage[hyperfootnotes=false]{hyperref}
\usepackage{xspace}
\usepackage{braket}
\usepackage{color}
\usepackage{units}
\usepackage{slashed} 
\usepackage{graphicx}
\usepackage{bm}

\newcommand{\be}{\begin{equation}}
\newcommand{\ee}{\end{equation}}
\newcommand{\bea}{\begin{eqnarray}}
\newcommand{\eea}{\end{eqnarray}}
\begin{document}

\title{$A_4$ realization of Linear Seesaw and Neutrino Phenomenology}

\author{M. Sruthilaya}
\email{msruthi28@gmail.com}
\affiliation{School of Physics,  University of Hyderabad, Hyderabad - 500046,  India }
\author{ Rukmani Mohanta}
\email{rmsp@uohyd.ernet.in}
\affiliation{School of Physics,  University of Hyderabad, Hyderabad - 500046,  India }
\author{Sudhanwa Patra}
\email{sudha.astro@gmail.com}
\affiliation{ Centre for Excellence in Theoretical and Mathematical Sciences
Siksha `O' Anusandhan University, Bhubaneshwar-751030, India}

\begin{abstract}
Motivated by the crucial role played by the discrete flavour symmetry groups in explaining the observed neutrino oscillation data,  we consider the $A_4$  realization of linear seesaw by extending the standard model (SM) particle content  
with two types of right-handed neutrinos along with the flavon fields, and  the SM symmetry with $A_4\times Z_4\times Z_2$ and a global symmetry $U(1)_X$ which is broken explicitly by the Higgs potential. We scrutinize whether this model can explain the recent results from neutrino oscillation experiments by searching for parameter space that can accommodate the observables such as the reactor mixing angle $\theta_{13}$, the CP violating phase  $\delta_{CP}$, sum of active neutrino masses $\Sigma_{i} m_i$, solar and atmospheric mass squared differences, 
and the lepton number violating parameter called as effective Majorana mass parameter, in line with recent experimental results. We also discuss  the scope of this model to explain the baryon asymmetry of the universe through Leptogenesis. 
We also investigate the possibility of probing the  non-unitarity effect in this scenario, but it is found to be rather small.
\end{abstract}
\maketitle 
 
\section{Introduction} 
The Standard Model (SM) of particle physics predicts massless neutrinos contradicting the experimental results on neutrino oscillation, according to which the three neutrino flavors mix with each other and at least two of the neutrinos have non-vanishing mass.  Due to the absence of right-handed (RH) neutrinos in the SM, neutrinos do not have Dirac mass like other charged fermions and their mass generation in the SM is generally expected to arise from a dimension-five operator \cite{Weinberg:1980bf}, which violates lepton number. However, very little is known about the origin of this operator and the underlying mechanism or its flavour structure. Hence, to  generate non-zero neutrino mass, one resorts to some beyond the standard model scenarios. There are many such models  where the SM is extended by  including the right-handed neutrinos to its particle content. The inclusion of right-handed neutrinos  $N_{R_i}$ not only generates the Dirac mass term but also leads to Majorana mass for the right handed neutrinos, which is of the form $\overline{N_{R_i}}N_{R_i}^c$ and violates $B-L$ symmetry. The smallness of active neutrino mass is ensured by the high value of Majorana mass of the right handed neutrino \cite{MINKOWSKI1977421,King:1998jw}. In these class of  models, if Dirac mass of neutrinos are of the order of lightest charged lepton mass i.e., electron mass, the Majorana mass has to be in TeV range  to get the observed value of active neutrino mass \cite{Mohapatra:1979ia}. But if such models have to be embedded in Grand Unified Theories (GUTs) where both Quarks and Leptons are treated on the same footing, the Dirac mass of neutrinos will be of the order of that of up-type quark \cite{Babu:1992hf} and the observed value of active neutrino mass requires Majorana mass to be of the order of $10^{15}$ GeV, which is beyond the access of present and future experiments. 

 Many possibilities were proposed to have not so heavy Majorana mass and the existence of other type of neutrinos called sterile neutrinos ($S$) is one among them \cite{Mohapatra:1986bd}. Now the neutrino mass can be expressed in the form of a $3\times 3$ matrix with each element represents a matrix. Depending on the position of the  zero elements in the mass matrix in the basis $(\nu, N_R, S)$, active neutrinos receive masses through two different mechanisms called inverse seesaw \cite{Mohapatra:1986bd,GonzalezGarcia:1988rw}  and linear seesaw \cite{Malinsky:2005bi}.  If 11 and 13  elements are zero then it is called inverse seesaw and if 11 and 33 elements are zero with non zero off-diagonal elements  then we have the linear seesaw. In all those cases the smallness of neutrino mass is independent of the ratio of Dirac mass to heavy neutrino mass, hence, allows to have heavy neutrinos in TeV range and bound on the ratio comes from non-unitarity effect and neutrinoless double beta decay experiments.
 
All those seesaw mechanisms require some of the elements of mass matrix to be zero or very small but none of them are prevented by SM symmetry. All those terms except 33 element in the matrix will be prohibited if the SM symmetry is extended to $SU(2)_L\times SU(2)_R\times SU(3)_C$ or $B-L$, since in those symmetry groups right handed neutrinos are no longer singlets. But linear seesaw requires 33 element of the mass matrix to be zero or very small which is difficult to obtain with gauge symmetry as sterile neutrinos are singlets in all gauge groups. But such terms will be absent if there is flavour symmetry under which sterile neutrinos have  non trivial representation.

The $A_4$ discrete symmetry group is the group of even permutations of  four elements has attracted a lot of attention since it is the smallest one which
admits one three-dimensional representation and three inequivalent one-dimensional representations.  Then, the
choice of the $A_4$ symmetry is natural since there are three families of fermions, i.e, the left-handed leptons can be
unified in triplet representation of $A_4$
while the right-handed leptons can be assigned to $A_4$
singlets.  This set-up was first proposed  in Ref. \cite{Ma:2001dn} to study the lepton masses and mixing obtaining nearly degenerate neutrino
masses and allowing realistic charged leptons masses after the
$A_4$ symmetry is spontaneously broken. Latter $A_4$ symmetry was proved to be very successful in generating Tribimaximal mixing pattern for Lepton mixing \cite{Altarelli:2005yp}, which well supported the trends of oscillation data at that time. The Tribimaximal mixing pattern  predicts solar mixing and atmospheric mixing angles consistent
with the experimental data but yields a vanishing reactor mixing angle \cite{Harrison:2002er,Harrison:2002kp,Harrison:2004uh} contradicting the recent experimental results
from the Daya Bay \cite{An:2012eh,An:2013uza}, T2K\cite{Abe:2013xua,Abe:2013hdq}, MINOS \cite{Evans:2013pka}, Double CHOOZ \cite{Crespo-Anadon:2014dea} and RENO \cite{Ahn:2012nd} experiments.  In view of
this the Tribimaximal mixing pattern has to be modified.

 Here we consider the realization of linear seesaw with $A_4$ symmetry. We extend SM symmetry with $A_4\times Z_4\times Z_2$  along with an extra global symmetry $U(1)_X$, as discussed in  Ref. \cite{Karmakar:2016cvb}. The SM particle content has been extended by introducing three RH neutrinos, $N_{R_i}$  and three  singlet fermions, $S_{R_i}$ along with the  flavon fields  ($\phi_S$, $\phi_T$, $\xi$, $\xi'$, $\rho$, $\rho'$), to understand the flavour structure of the lepton mixing. The proposed model 
gives almost similar result  as in   \cite{Karmakar:2016cvb} in the context of neutrino oscillation, but has a different physics aspect in the case of heavy neutrinos. In  \cite{Karmakar:2016cvb}, the active neutrinos get their mass through inverse seesaw with the prediction of six nearly degenerate heavy neutrinos but in our case there are three very different mass state with each state is nearly doubly degenerated.  Also, our proposed scenario is very much suitable for Leptogenesis  as discussed in \cite{Gu:2010xc, Pilaftsis:1997jf}, where the analytic expression for CP asymmetry and corresponding baryon asymmetry for the case of three pairs of nearly degenerate heavy neutrinos can be found.  In Ref. \cite{Pilaftsis:1997jf},  the contributions of the absorptive part of Higgs self-energy to CP violation in heavy particle decays termed as $\epsilon$-type CP violation, has been discussed elaborately. Such contributions are neglected in many cases as they are small compared to $\epsilon^{\prime}$-type, the CP violation in heavy neutrino decays due to the overlapping of tree-level with one-loop vertex diagram. They have provided the formalism to deal with mixing of states during the decay of the particles and have shown that there is resonant enhancement of $\epsilon$-type CP violation, if mixing states are nearly degenerate. The CP asymmetries due to both types of CP violations for a model with a pair of nearly degenerate heavy neutrinos are also calculated and it was shown that the  CP asymmetry due to $\epsilon$-type CP violation is 100 times  more than that of due to $\epsilon^{\prime}$-type, which in turn   predicts  the correct baryon asymmetry of the Universe. 

The outline of the paper is as follows. In section II, we present the model framework for linear seesaw.   The $A_4$ realization of linear seesaw and its implication to neutrino oscillation parameters is discussed in Section III.
Section IV contains the discussion on Leptogenesis and summary and Conclusions are presented in Section V.
 
\section{The model framework for linear seesaw}
We consider the minimal extension of Standard Model $\mathcal{G}_{\rm SM} \equiv SU(2)_L \times U(1)_Y$,
omitting the $SU(3)_C$ structure for simplicity,
with two types of singlet neutrinos which are complete singlet under $\mathcal{G}_{\rm SM}$ for implementation 
of linear seesaw. We denote these neutral fermion singlets as right-handed sterile neutrinos $N_{R_i}$ 
and  $S_{R_i}$. Both these neutral fermion species have Yukawa coupling 
with the lepton doublet $L$. In addition, one can write down a mixing term connecting these 
two species of neutrinos. The bare Majorana mass terms for $N_{R_i}$ and $S_{R_i}$ are either assumed 
to be zero or forbidden by  some symmetry arguments. The leptonic Lagrangian for linear seesaw 
mechanism is given by
\begin{eqnarray}
	-\mathcal{L} &=&
	y \overline{L} \tilde{H} N_\text{R} + 
	h \overline{L} \tilde{H} S_\text{R} + 
	\overline{N_R} m_{RS} S^c_\text{R} + 
	\text{~h.c.}  \nonumber \\
	&=&
	\overline{\nu}_\text{L} m_\text{D} N_\text{R} + 
	\overline{\nu}_\text{L} m_\text{LS} S_\text{R} + 
	\overline{N}_\text{R} m_{RS} S^c_\text{R} + 
	\text{~h.c.}\;.
\label{model_lagrangian}
\end{eqnarray}
The full mass matrix for neutral leptons in the basis 
$N = (\nu_\text{L}, ~N^c_\text{R}, ~S^c_\text{R})^\text{T}$
is given as
\begin{equation}
	\mathbb{M} = 
	\begin{pmatrix}
		0 & m_\text{D} & m_\text{LS} \\
		m_\text{D}^\text{T} & 0 & m_{RS} \\
		m_\text{LS}^\text{T} & m^\text{T}_{RS} & 0
	\end{pmatrix}.		
\end{equation}
The resulting mass formula for light neutrinos is governed by linear seesaw mechanism,
\begin{align}
m_\nu &= m_\text{D} m_{RS}^{-1} m_\text{LS}^\text{T} \mbox{+ transpose}\;.
\label{numass_linear}
\end{align}

\section{An $A_4$ realization of linear seesaw}
In this section, we wish to present an $A_4$ realization of linear seesaw which has 
been discussed in the previous section. The particle content of the model and their representations 
under flavour symmetries are presented in Table\,\ref{table:t1}\footnote{The implication of linear 
seesaw can be found in \cite{Gu:2010xc}.}. 
We introduce an extra global symmetry $U(1)_X$ which is broken explicitly but softly by the term $\mu^2_{\rho\xi}\rho\xi+{\rm h.c.}$, in the Higgs potential to prevent Goldstone boson \cite{He:2006dk}. This term not only breaks $U(1)_X$ symmetry but also gives non-zero vacuum expectation value to $\rho$, $\langle \rho \rangle =\frac{\mu^2_{\rho\xi}\langle \xi \rangle}{m_{\rho}^2}\ll \langle \xi\rangle $ as $\mu^2_{\rho\xi}$ is very small compared to $m_{\rho}$, the mass of $\rho$.

\begin{table}[h]
\begin{center}
\begin{small}
\vspace*{0.1 true in}
\begin{tabular}{|c|ccccccc|cccccc|}
\hline
~Fields~ &~~~$e_R$~ &~ $\mu_R$~ &~ $\tau_R$ ~ &  ~$L$ ~ &~  $H$~  & 
 ~$N_R$~ & ~$S_R$~  &~ $\phi_T$~  & ~ $\phi_S$~  &~ $\xi$~  & ~$\xi^{\prime}$~ & ~$\rho$~ &~ $\rho^{\prime}$~~\tabularnewline
\hline
$A_4 $ & 1 & $1^{\prime\prime}$ & $1^{\prime}$ &3 &1 &3 &3 &3 &3  &1 &$1^{\prime}$ &1 &1 \tabularnewline
\hline 

$Z_4 $ & $-i$ & $-i$ &$-i$ &$-i$ &1 &$i$  &1 & 1 &$i$ &$i$ &$i$ &$-i$ &$-1$\tabularnewline
\hline

$Z_3 $ & 1  & 1 &1 &1 &1 &1 &$\omega$ &1 &$\omega$ &$\omega$ &$\omega$ &$\omega^2$ &1\tabularnewline
\hline
$X $ & $-1$  & $-1$ & $-1$& $-1$&0 &0 &0 &0 &0 &0 &0 &$-1$&$-1$\tabularnewline
\hline
\end{tabular}
\caption{The particle content and their charge assignments for an $A_4$ realization of 
         linear seesaw mechanism.}
\label{table:t1}
\end{small}
\end{center}
\end{table}
The Yukawa Lagrangian for  the charged  lepton sector is  given as 
\begin{eqnarray}\nonumber
&&\mathcal{L}_l =-\left\{
\left[ \frac{\lambda_e}{\Lambda} \left( \bar{L}\phi_T\right)H e_{R}\right]+
\left[ \frac{\lambda_{\mu}}{\Lambda}\left(\bar{L}\phi_T\right)^{\prime}H\mu_{R}\right]+
\left[ \frac{\lambda_{\tau}}{\Lambda}\left(\bar{L}\phi_T\right)^{\prime\prime}H\tau_{R}\right] 
\right\}\,.\nonumber
\end{eqnarray}
After giving non-zero VEVs to SM Higgs as well as flavon fields and breaking all symmetries, 
the mass matrix for charged leptons is found to be
\begin{equation}
M_l=v\frac{v_T}{\Lambda}{\rm diag}\left(\lambda_e,\lambda_{\mu},\lambda_{\tau}\right),
\end{equation}
where  the vacuum expectation values (vevs) of the scalar fields are given as
\begin{equation}
v=\langle H \rangle,~~v_T=\langle \phi_T\rangle\;.
\end{equation}

For linear seesaw mechanism, the Lagrangian involved in the generation of the mass matrices for 
an $A_4$ flavor symmetric framework can be written as,
\begin{align}\label{a4lag}
 -\mathcal{L_{\nu}}=&\mathcal{L}_{\nu N}+\mathcal{L}_{N S}+\mathcal{L}_{\nu S}\;,
\end{align}
where
\begin{align}
\mathcal{L}_{\nu N} & = y_1 \overline{L}\widetilde{H} N_R \frac{\rho^{\prime}}{\Lambda}\;,  \\
\mathcal{L}_{\nu S} & = y_2 \overline{L} \widetilde{H} S_R\frac{\rho}{\Lambda}\;,           \\
\mathcal{L}_{N S} & =   \left(\lambda_{NS}^{\phi}\phi_s+\lambda_{NS}^{\xi}\xi+
                          \lambda_{NS}^{\xi^{\prime}}\xi^{\prime}\right) \overline{N_R} S^c_R\;.
\end{align}
It should be noted that the terms $\mathcal{L}_{\nu N}$, $\mathcal{L}_{\nu S}$ and $\mathcal{L}_{N S}$ represent the  contributions 
for Dirac neutrino mass connecting $\nu_L-N_R$, $\nu_L-S_R$  mixing and $N_R-S^c_R$ mixing terms. 
If one looks at the mass formula for light neutrinos governed by linear seesaw mechanism  given in Eq. (\ref{numass_linear}), one can use the mass hierarchy as $m_{RS} \gg m_D, m_{LS}$. That is the 
reason why we forbid $\overline{\nu} N$ and $\overline{\nu} S$ terms at tree level and generate them by dimension five operator while the heavy mixing term $N-S$ is generated at tree level. 

Using the   following vevs for the scalar and flavon fields
\begin{eqnarray}\label{e4a}
\langle \phi_S \rangle= v_S(1,1,1),~~ \langle\xi \rangle=v_{\xi},~ \langle\xi^{\prime}\rangle
=v_{\xi^{\prime}},~~
\langle\rho \rangle=v_{\rho},~~ \langle\rho^{\prime}\rangle=v_{\rho^{\prime}}\;,  \nonumber
\end{eqnarray}
the various mass matrices are found to be
\begin{eqnarray}
&&m_\text{D}=y_1v\frac{v_{\rho^{\prime}}}{\Lambda}\left(
\begin{array}{ccc}
1 & 0 & 0\\
0 & 0 &1 \\
0 & 1 &0
\end{array}
\right)
,~~~~m_\text{LS}=y_2v\frac{v_{\rho}}{\Lambda}
\left(
\begin{array}{ccc}
1 & 0 & 0\\
0 & 0 &1 \\
0 & 1 &0
\end{array}
\right),\\ \label{e4-a}
&&m_{RS}=\frac{a}{3}\left(
\begin{array}{ccc}
2 & -1 & -1\\
-1 & 2 &-1 \\
-1 & -1 &2
\end{array}
\right)+
b\left(
\begin{array}{ccc}
1 & 0& 0\\
0 & 0 &1 \\
0& 1 &0
\end{array}
\right)+
d\left(
\begin{array}{ccc}
0 & 0& 1\\
0 & 1 &0 \\
1& 0 &0
\end{array}
\right),
\end{eqnarray}
where $a=\lambda_{NS}^{\phi}v_S$, $b=\lambda_{NS}^{\xi}v_{\xi}$ and $d=\lambda_{NS}^{\xi^{\prime}}v_{\xi^{\prime}}$. 

 The first term in  Eqn. (\ref{e4-a}) comes from $\lambda_1\phi_s\left(\overline{N_R}S_R^c\right)_{3s}$, where $\left(\overline{N_R}S_R^c\right)_{3s}$ is a triplet which is symmetric under exchange of $N_R$ and $S_R$. The product of two triplets can also form a triplet which is antisymmetric under the exchange of the particles. In linear seesaw, the mass of the light neutrino  is represented as $m_{\nu}=m_\text{D}m_\text{RS}^{-1}m_\text{LS}^{T}$+transpose, and as seen from Eqn (\ref{e4a}) the mass matrices  $m_\text{D}$ and $m_\text{LS}$ are symmetric and are related as  $m_\text{D} \propto m_\text{LS}$. Hence, in $m_{\nu}=m_\text{D}^{T}(m_\text{RS}^{-1}+{m_\text{RS}^{-1}}^T)m_\text{LS}$, the antisymmetric part cancels out and only symmetric part survives. 

\section{Neutrino Masses and Mixing}
\label{sec:mass-mixing}
For calculational convenience one can rewrite the $m_\text{RS}$ mass matrix
(\ref{e4-a}) as
\begin{equation}
m_{RS}=\left(
\begin{array}{ccc}
2a/3+b & -a/3& -a/3\\
-a/3 & 2a/3 &-a/3+b \\
-a/3& -a/3+b &2a/3
\end{array}
\right)+
\left(
\begin{array}{ccc}
0 & 0& d\\
0 & d &0 \\
d& 0 &0
\end{array}
\right).
\end{equation}
Thus, with Eqns. (\ref{numass_linear}), (\ref{e4a}) and (\ref{e4-a}),
 one can obtain the the light neutrino mass
\begin{eqnarray}
m_{\nu}&=&m_\text{D} m_{RS}^{-1}m_\text{LS}^T+\text{transpose} \nonumber\\
&=&k_1k_2\left(\begin{array}{ccc}
1 & 0 & 0\\
0 & 0 &1 \\
0 & 1 &0
\end{array}\right) m_{RS}^{-1}\left(\begin{array}{ccc}
1 & 0 & 0\\
0 & 0 &1 \\
0 & 1 &0
\end{array}\right),
\end{eqnarray}
where the parameters $k_1$ and $k_2$  are related to the vevs through
\begin{eqnarray}
k_1=\sqrt{2} y_1v\frac{v_{\rho^{\prime}}}{\Lambda}\;, ~~~~k_2=\sqrt{2}y_2v\frac{v_{\rho}}{\Lambda}.\nonumber
\end{eqnarray}
 Hence, the inverse of light neutrino mass matrix is given by
\begin{eqnarray}
m_{\nu}^{-1}
&=&\frac{1}{k_1k_2}\left(
\begin{array}{ccc}
2a/3+b & -a/3& -a/3\\
-a/3 & 2a/3 &-a/3+b \\
-a/3& -a/3+b &2a/3
\end{array}
\right)+\frac{1}{k_1k_2}\left(
\begin{array}{ccc}
0 & d& 0\\
d & 0 &0 \\
0& 0 &d
\end{array}
\right),
\end{eqnarray}
which in TBM basis will have the form, i.e., $m_{\nu}^{-1^{\prime}}=U_\text{TBM}^T m_{\nu}^{-1}U_\text{TBM}$, 
\begin{equation}
m_{\nu}^{-1^{\prime}}=\left(\begin{array}{ccc}
a+b-d/2 & 0 & -\frac{\sqrt{3}}{2}d\\
0 &~~ b+d~~ &0 \\
-\frac{\sqrt{3}}{2}d & 1 &a-b+d/2
\end{array}\right)\;.
\end{equation}
The inverse mass matrix $m_{\nu}^{-1^{\prime}}$ can be  diagonalized by $U_{13}^*$. Hence, the  matrix $m_\nu^{-1}$  can be diagonalized by $U_{TBM}\cdot U_{13}^*$, and thus, the matrix $m_{\nu}$ can be diagonalized  by $U_{TBM}\cdot U_{13}$,  while $m_{RS}$  by $U_{TBM}\cdot U_{13}^T$.  The complex unitary matrix $U_{13}$ has the form 
 \begin{equation}
 U_{13}=\left(
 \begin{array}{ccc}
 \cos\theta & 0 & \sin\theta e^{-i\psi}\\
 0 & 1 & 0 \\
 -\sin\theta e^{i\psi}& 0 & \cos\theta
 \end{array}
 \right),
 \end{equation}
 where the parameters $\theta$ and $\psi$ are expressed in terms of the 
 mass matrix parameters  $d/b=\lambda_1e^{\phi_{db}}$, $a/b=\lambda_2e^{\phi_{ab}}$ as
 \begin{eqnarray}
 \tan 2\theta=-\frac{\sqrt{3}\lambda_1\cos\phi_{db}}{(\lambda_1 \cos\phi_{db}-2)\cos\psi +(2\lambda_2\sin\phi_{ab})\sin\psi}\;,\label{tant}
 \end{eqnarray}
 and
 \begin{eqnarray}
 \tan\psi=\frac{\sin\phi_{db}}{\lambda_2\cos(\phi_{ab}-\phi_{db})}\;.\label{tanpsi}
 \end{eqnarray}
 The eigenvalues of $m_{\nu}$ and $m_{RS}$ are related to each other as
 \begin{equation}
 \tilde{m}_i=\frac{k_1k_2}{\tilde {M}_i}\;.\label{e16}
 \end{equation}
  where $\tilde{m}_i$ and $\tilde{M}_i$ are $i^\text{th}$ eigenvalues of $m_{\nu}$ and $m_{RS}$ respectively. The eigenvalues of $m_{RS}$  can be expressed as 
  \begin{eqnarray}
\tilde{M}_1&=&b\left[\lambda_2 e^{i\phi_{ab}}-\sqrt{1+\lambda_1^2e^{2i\phi_{db}}-\lambda_1 e^{i\phi_{db}}}\right], \nonumber\\
\tilde{M}_2&=&b\left[1+\lambda_1 e^{i\phi_{db}}\right], \nonumber\\
\tilde{ M}_3&=&b\left[\lambda_2 e^{i\phi_{ab}}+\sqrt{1+\lambda_1^2e^{2i\phi_{db}}-\lambda_1 e^{i\phi_{db}}}\right],\label{eq17a} 
 \end{eqnarray}
 which give the mass of the heavy neutrinos as $M_i=|\tilde{M_i}|$.
Explicitly, one can  write the heavy neutrino masses as
 \begin{eqnarray}
M_1&=& |b|M_1^{\prime}=|b|\left[(\lambda_2\cos\phi_{ab}-C)^2+(\lambda_2\sin\phi_{ab}-D)^2\right]^{1/2} \nonumber\\
 M_2&=&|b|M_2^{\prime}=|b|\left[1+\lambda_1^2+2\lambda_1\cos\phi_{db}\right]^{1/2} \nonumber\\
 M_3&=&|b|M_3^{\prime}=|b|\left[(\lambda_2\cos\phi_{ab}+C)^2+(\lambda_2\sin\phi_{ab}+D)^2\right]^{1/2}\;,\label{eq17} 
 \end{eqnarray}
 where
 \begin{eqnarray}
 C&=&\pm\left[ \frac{A+\sqrt{A^2+B^2}}{2}\right]^{1/2}\;,~~~~~~~
 D=\pm\left[ \frac{-A+\sqrt{A^2+B^2}}{2}\right]^{1/2}\;,\nonumber\\
 A&=&1+\lambda_1^2\cos 2\phi_{db}-\lambda_1\cos\phi_{db}\;,~~~~
 B=\lambda_1^2\sin 2\phi_{db}-\lambda\sin\phi_{db}\;.\label{abcd}
 \end{eqnarray}
 and the phases $\phi_i$s as
 \begin{eqnarray}
 \phi_1&=&\tan^{-1}\left[\frac{\lambda_2\sin\phi_{ab}-D}{\lambda_2\sin\phi_{ab}-C}\right]\;,\nonumber \\
 \phi_2&=&\tan^{-1}\left[\frac{\lambda_1\sin\phi_{db}}{1+\lambda_1\cos\phi_{db}}\right]\;, \nonumber\\
  \phi_3&=&\tan^{-1}\left[\frac{\lambda_2\sin\phi_{ab}+D}{\lambda_2\sin\phi_{ab}+C}\right].
 \end{eqnarray}
 Thus, the active neutrino masses $m_i=|\tilde{m}_i|$ and the matrix which diagonalize active neutrino mass matrix, $U_{\nu}$ are given by
 \begin{eqnarray}
 m_i&=&\frac{|k_1k_2|}{M_i}\;,\nonumber\\
 U_{\nu}&=&U_{TBM}\cdot U_{13}\cdot P\;,
\end{eqnarray}  
with $P=\text{diag}(e^{-i\phi_1/2},e^{-i\phi_2/2},e^{-i\phi_3/2})$.
\\
 The lepton mixing matrix, known as PMNS matrix is given by \cite{Pontecorvo:1957qd,Maki:1962mu} 
\begin{equation}
U_{PMNS}=U_l^{\dagger}\cdot U_{\nu}\;,
\end{equation}
where $U_l$ and $U_{\nu}$ are the matrices which diagonalize charged lepton and neutrino mass matrices. Here $U_l=I$ and $U_{\nu}=U_{TBM}\cdot U_{13}\cdot P$, hence,
\begin{equation}
U_{PMNS}=U_{TBM}\cdot U_{13}\cdot P, \label{eqn15}
\end{equation} 
which is proved to be in good agreement with the experimental observations \cite{Chao:2011sp,M.:2014kca}. The PMNS matrix can be parametrized in terms of three mixing angles ($\theta_{13}$, $\theta_{23}$ and $\theta_{12}$) and three phases (one Dirac phase $\delta_{CP}$, and two Majorana phases $\rho$ and $\sigma$) as
\begin{equation}
U_\text{PMNS}=\left( \begin{array}{ccc} c^{}_{12} c^{}_{13} & s^{}_{12}
c^{}_{13} & s^{}_{13} e^{-i\delta_{CP}} \\ -s^{}_{12} c^{}_{23} -
c^{}_{12} s^{}_{13} s^{}_{23} e^{i\delta_{CP}} & c^{}_{12} c^{}_{23} -
s^{}_{12} s^{}_{13} s^{}_{23} e^{i\delta_{CP}} & c^{}_{13} s^{}_{23} \\
s^{}_{12} s^{}_{23} - c^{}_{12} s^{}_{13} c^{}_{23} e^{i\delta_{CP}} &
-c^{}_{12} s^{}_{23} - s^{}_{12} s^{}_{13} c^{}_{23} e^{i\delta_{CP}} &
c^{}_{13} c^{}_{23} \end{array} \right) P^{}_\nu \;,\label{pmns}
\end{equation}
where $c_{ij}=\cos\theta_{ij}$ and $s_{ij}=\sin\theta_{ij}$ and $P_{\nu}=\text{diag}(e^{i\rho},e^{i\sigma},1)$. 
From Eqns. (\ref{eqn15}) and (\ref{pmns}), one can find 
\begin{eqnarray}
&&\sin\theta =\sqrt{\frac{3}{2}}\sin\theta_{13} \nonumber \\
&&\sin\delta_\text{CP}=-\frac{\sin\psi}{\displaystyle{\sqrt{1-\frac{3(2-3\sin^2\theta_{13})}{(1-\sin^2\theta_{13})^2}\sin^2\theta_{13}\cos^2\psi}}}\approx -\sin\psi\;.
\end{eqnarray}
  The above expressions relate the parameters of the model, i.e.,  $\theta$ and $\psi$ to the mixing observables $\sin^2\theta_{13}$ and $\delta_{CP}$ respectively. Since $\sin^2\theta_{13}$ is known more precisely than $\delta_\text{CP}$, in our calculation we fix $\theta$ by fixing $\sin^2\theta_{13}$ at its best fit value while considering all possible value of $\psi$ for which $\delta_\text{CP}$ falls within its $3\sigma$ experimental range.\\
 
\section{Numerical results}
 
 Using Eqns. (\ref{e16}) and (\ref{eq17}), the light neutrino masses are found to be 
 \begin{equation}
  m_i=\frac{|k_1k_2|}{M_i}=\frac{|k_1k_2|}{|b|}\frac{1}{M_i^{\prime}}\;.\label{lmass}
\end{equation}
 Since  only the mass squared differences, $\Delta m^2_{21}$ (solar mass squared difference) and $|\Delta m^2_{32}|$ (atmospheric mass squared difference) are measured in neutrino oscillation experiments, we calculate the mass squared differences from Eqn. (\ref{lmass}) as
\begin{eqnarray}
\Delta m^2_{21}&=&\left|\frac{k_1k_2}{b}\right|^2\left(\frac{1}{{M_2^{\prime}}^2}-\frac{1}{{M_1^{\prime}}^2}\right)\;, \nonumber \\
 \left|\Delta m^2_{31}\right|&=&\left|\frac{k_1k_2}{b}\right|^2\left|\left(\frac{1}{{M_3^{\prime}}^2}-\frac{1}{{M_1^{\prime}}^2}\right)\right|\;. \label{mdiff}
\end{eqnarray}
Substituting the set of Eqns. (\ref{eq17}) in the above equations, we find
 the ratio of  the two mass squared differences as 
 \begin{eqnarray}
 r&=&\frac{\Delta m_{21}^2}{|\Delta m_{31}^2|}=\left[\frac{(\lambda_2\cos\phi_{ab}+C)^2+(\lambda_2\sin\phi_{ab}+D)^2}{1+\lambda_1^2+2\lambda_1\cos\phi_{db}}\right] \nonumber\\
 &\times &\left[\frac{(\lambda_2\cos\phi_{ab}-C)^2+(\lambda_2\sin\phi_{ab}-D)^2-\left(1+\lambda_1^2+2\lambda_1\cos\phi_{db}\right)}{4\lambda_2|C \cos\phi_{ab}+D\sin\phi_{ab}|}\right]\;.\label{ra}
 \end{eqnarray}
Now using equations (\ref{tant}), (\ref{tanpsi}), (\ref{eq17}), (\ref{abcd}) and \ref{ra}, and by fixing the parameters $\phi_{db}$, $\psi$ and $\theta$, one can find numerical values of $M_i'$'s. Once $M_i'$'s are known $\left|\displaystyle{\frac{k_1k_2}{b}}\right|$ can be calculated from (\ref{mdiff}) as
\begin{equation}
\left|\frac{k_1k_2}{b}\right|=\sqrt{\frac{\Delta m^2_{21}}{\displaystyle{ \left (\frac{1}{M_2^{\prime 2}}-\frac{1}{M_1^{\prime 2}}\right )}}}
 =\sqrt{\left|\frac{\Delta m^2_{31}}{\left (\frac{1}{M_3^{\prime 2}}-\frac{1}{M_1^{\prime 2}}\right )}\right|}\;,
\end{equation}
which will also give the absolute value of light neutrino masses as all the quantities on the right hand side of (\ref{lmass}) are now known.

 We now rewrite the expression  $\tan \psi$ (\ref{tanpsi})  in terms of $\phi_{db}$ as
\begin{equation}
\phi_{db}=0,\pi,~~~\text{for}~ \tan\psi=0\;,
\end{equation}
and 
\begin{equation}
\phi_{ab}=\phi_{db}+ \cos^{-1}\left(\frac{\sin\phi_{db}}{\lambda_2\tan\psi}\right),~~~\text{for}~~ \tan\psi\neq 0,
\end{equation}
 and consider the following cases to see the implications.\\

\begin{figure}[!t]
\includegraphics[width=6cm,height=5.0cm]{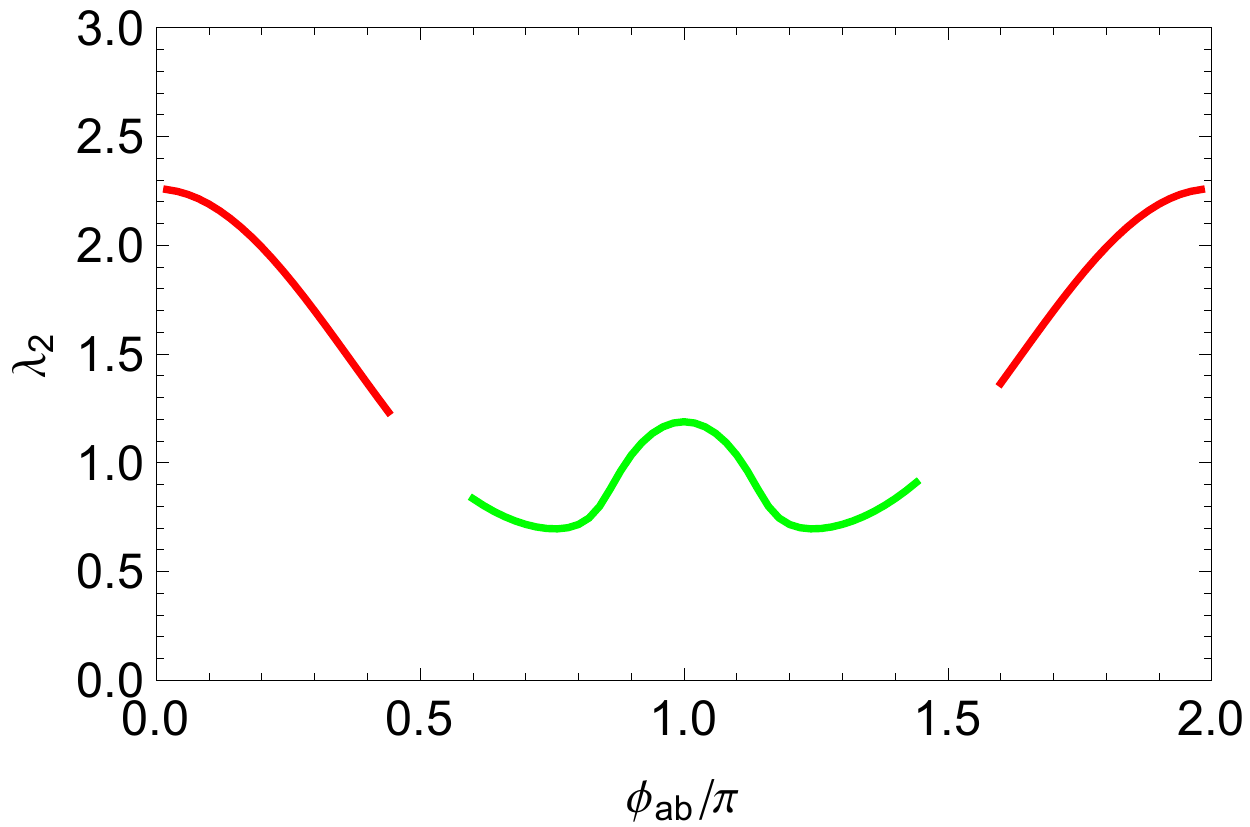}
\hspace{0.2 truein}
\includegraphics[width=6cm,height=5.0cm]{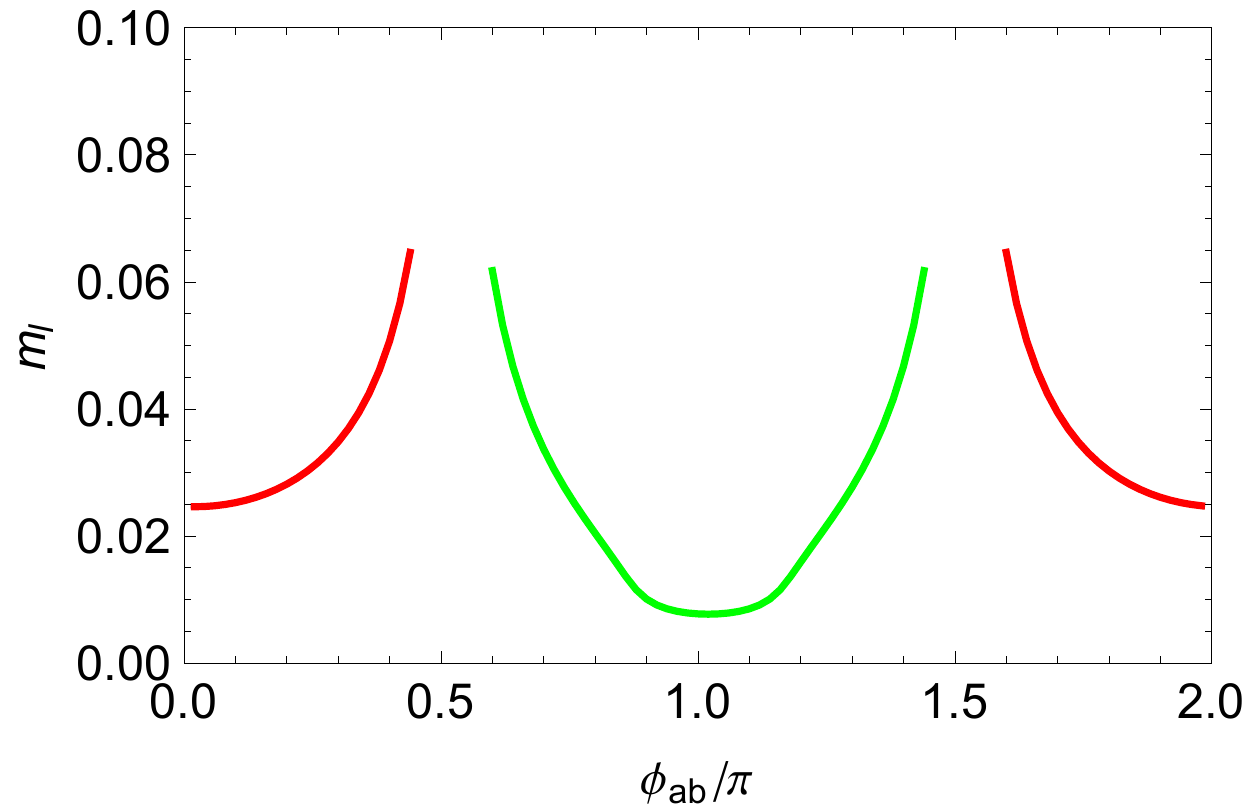}\\
\vspace{0.2 truein}
\includegraphics[width=6cm,height=5.0cm]{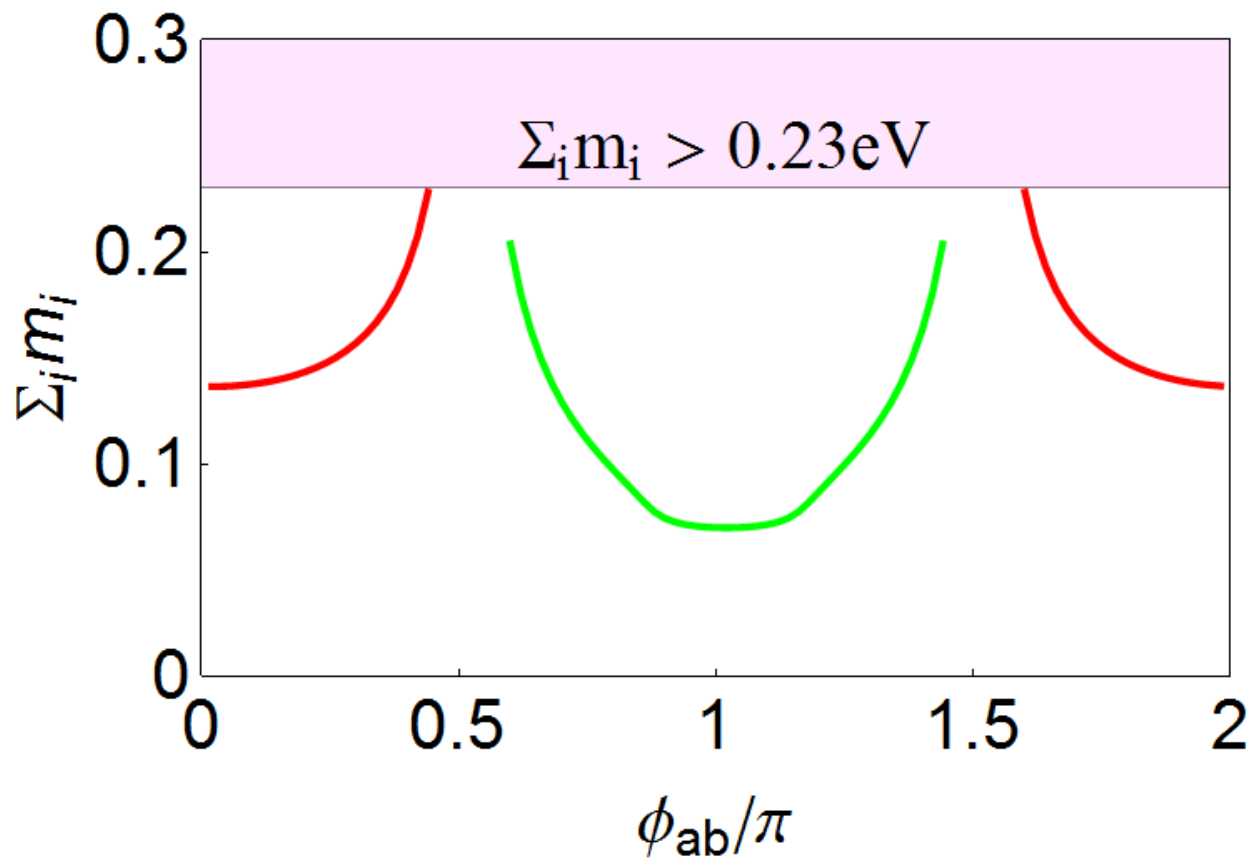}
\caption{Variation of $\lambda_2$, the lightest neutrino mass ($m_l$) and $\Sigma_im_i$ with $\phi_{ab}$, red lines are for inverted hierarchy and green lines are for normal hierarchy. }\label{c1a}
\end{figure}

\subsection{Correlation between model parameters with $\tan\psi=0$}
 
 In this case  $\phi_{db}$ will be either $0$ or $\pi$, and for $\phi_{db}=0$ one can obtain from Eq. (\ref{tant}) 
 \begin{equation} 
 \lambda_1=\frac{2\tan2\theta}{\sqrt{3}+\tan2\theta}\;,
 \end{equation}
 and
 the ratio of the mass square differences $r$ satisfies the relation 
 \begin{equation}
 r=0.03=\left[\frac{\lambda_2^2+2\lambda_2C\cos\phi_{ab}+C^2}{(1+\lambda_1)^2}\right]\left[\frac{\lambda_2^2-2\lambda_2C\cos\phi_{ab}+C^2-(1+\lambda_1)^2}{4\lambda_2|C\cos\phi_{ab}|}\right], \label{rb}
 \end{equation}
 where $C=\sqrt{\frac{1-\lambda_1+\lambda_1^2}{2}}$. 
  The eigenvalues of $m_{RS}$ in this case become
 \begin{eqnarray}
 M_1&=&|b|\sqrt{\lambda_2^2-2\lambda_2C\cos\phi_{ab}+C^2}\;, \nonumber\\
 M_2&=&|b|(1+\lambda_1)\;,\nonumber\\  
 M_3&=&|b|\sqrt{\lambda_2^2+2\lambda_2C\cos\phi_{ab}+C^2}\;.
 \end{eqnarray}
 Now from Eq. (\ref{rb}), using the measured values of $r~(0.03)$, variation of the  parameter $\lambda_2$, the lightest neutrino mass ($m_l$) and the sum of active neutrino masses $\sum m_i$  with $\phi_{ab}$ are shown in Fig.\ref{c1a}.
 
\begin{figure}[!t]
\includegraphics[width=6cm,height=5.0cm]{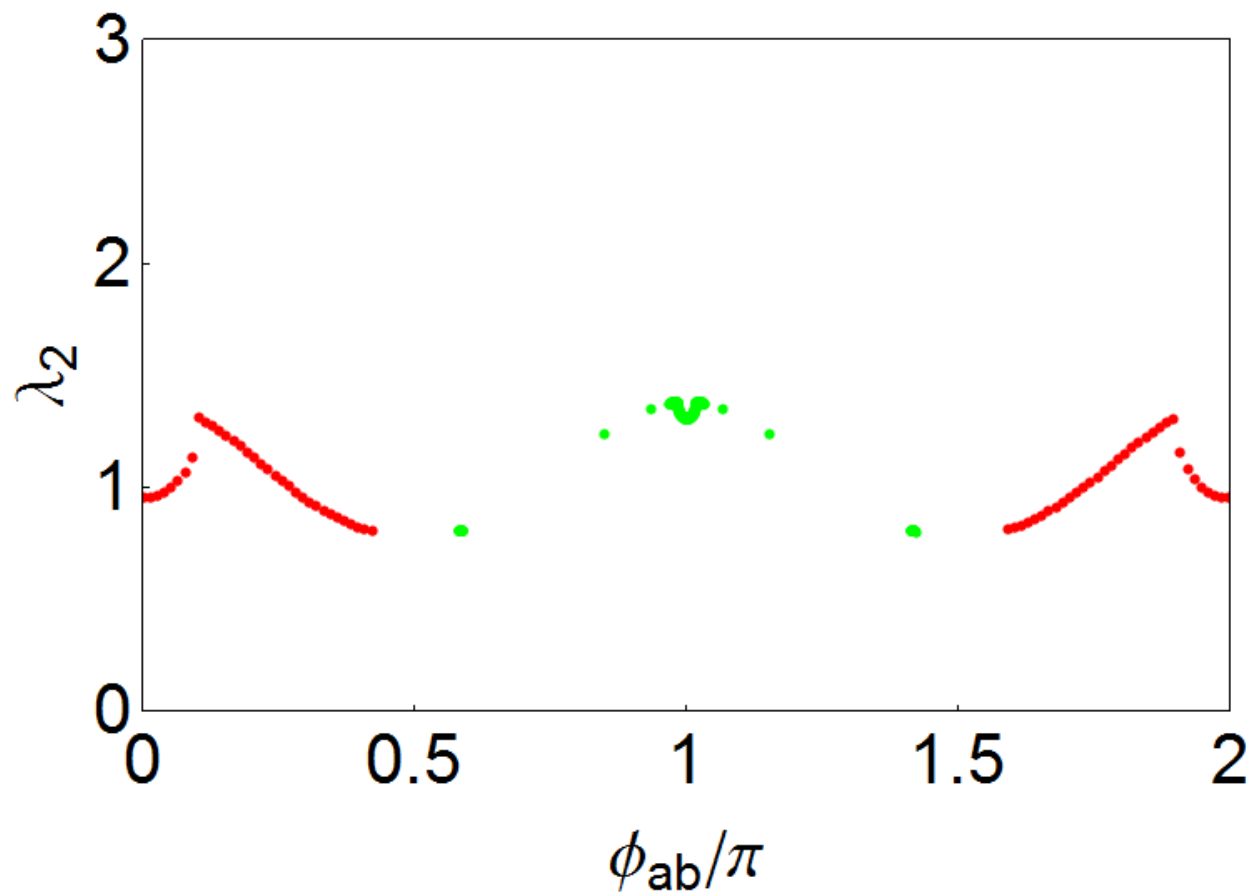}
\hspace{0.2 truein}
\includegraphics[width=6cm,height=5.0cm]{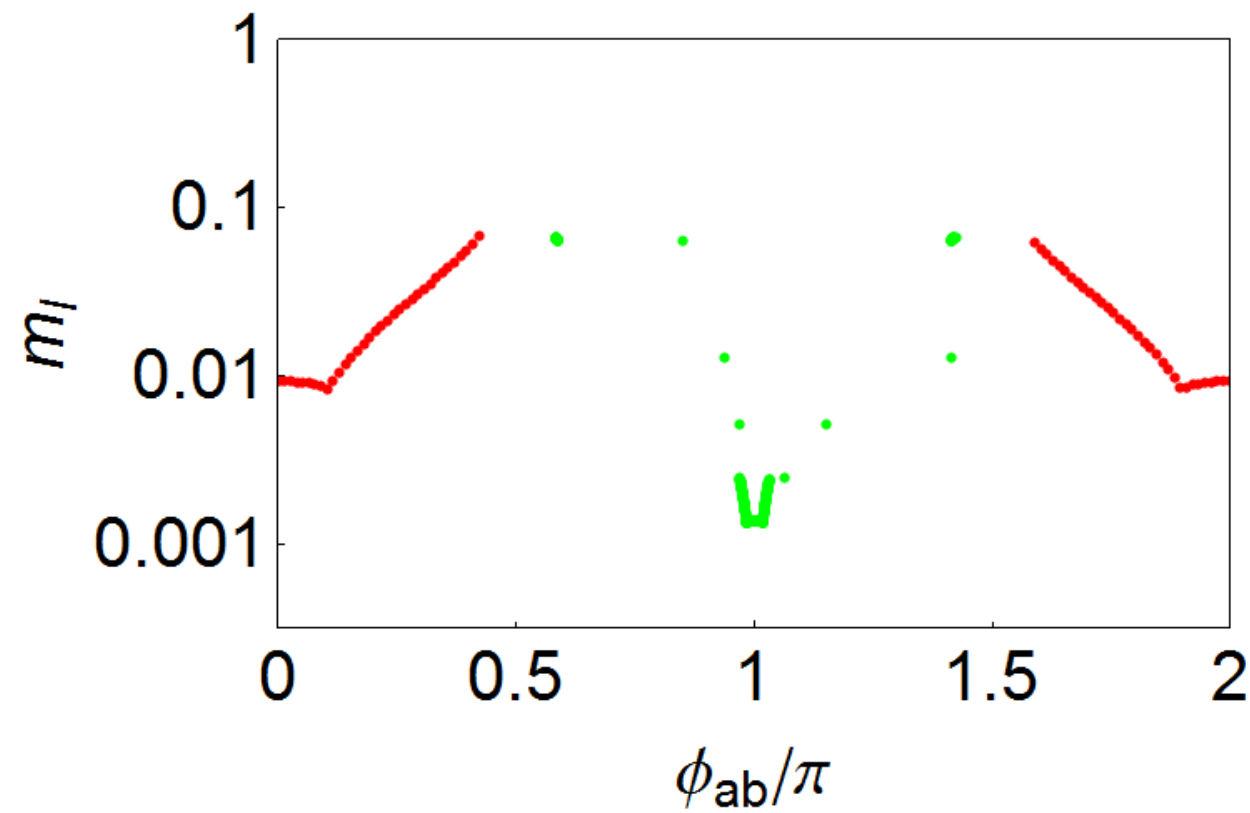}\\
\vspace{0.2 truein}
\includegraphics[width=6cm,height=5.0cm]{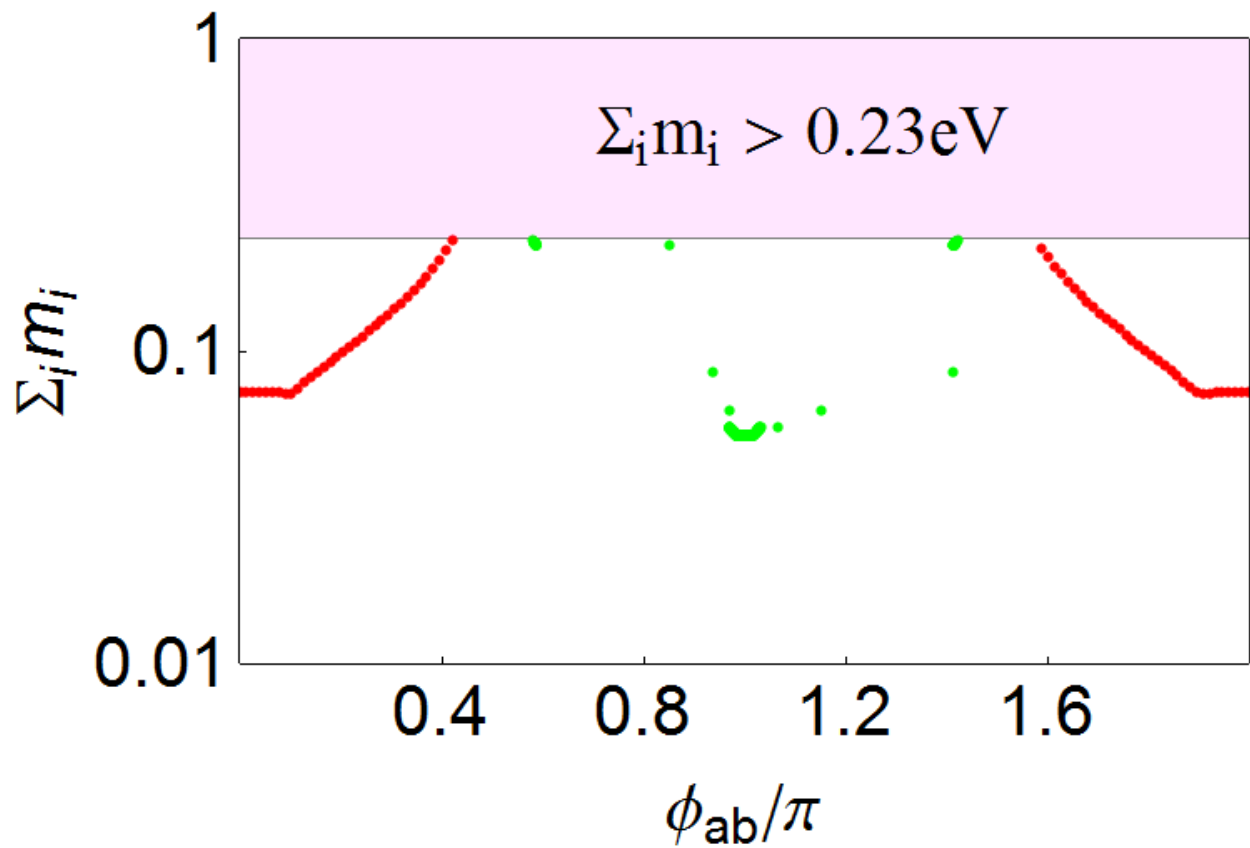}
\caption{Variation of $\lambda_2$, $m_l$ and $\Sigma_im_i$ with $\phi_{ab}$, red points are for inverted hierarchy and green points  are for normal hierarchy. }\label{f1}
\end{figure}

While for $\phi_{db}=\pi$
\begin{equation} 
 \lambda_1=\frac{2\tan2\theta}{\sqrt{3}-\tan2\theta}\;,
 \end{equation}
 and
 the ratio of the mass square differences $r$ satisfies the relation 
 \begin{equation}
 r=0.03=\left[\frac{\lambda_2^2+2\lambda_2C\cos\phi_{ab}+C^2}{(1-\lambda_1)^2}\right]\times\left[\frac{\lambda_2^2-2\lambda_2C\cos\phi_{ab}+C^2
 -(1-\lambda_1)^2}{4\lambda_2|C\cos\phi_{ab}|}\right], \label{r}
 \end{equation}
  with $C=\sqrt{\frac{1+\lambda_1+\lambda_1^2}{2}}$,
and  
 the eigenvalues of $m_{RS}$ are given as 
 \begin{eqnarray}
 M_1&=&|b|\sqrt{\lambda_2^2-2\lambda_2C\cos\phi_{ab}+C^2}\;, \nonumber\\
 M_2&=&|b|(1-\lambda_1)\;,\nonumber\\  
 M_3&=&|b|\sqrt{\lambda_2^2+2\lambda_2C\cos\phi_{ab}+C^2}\;.
 \end{eqnarray}
Analogous to Fig.1, the variation of various parameters with $\phi_{ab}$ is shown in Fig.\ref{f1}. From the plots it can be seen that for normal ordering, the allowed parameter space is severely constrained.
 \begin{figure}[!t]
 \includegraphics[width=6.0cm,height=5.0cm]{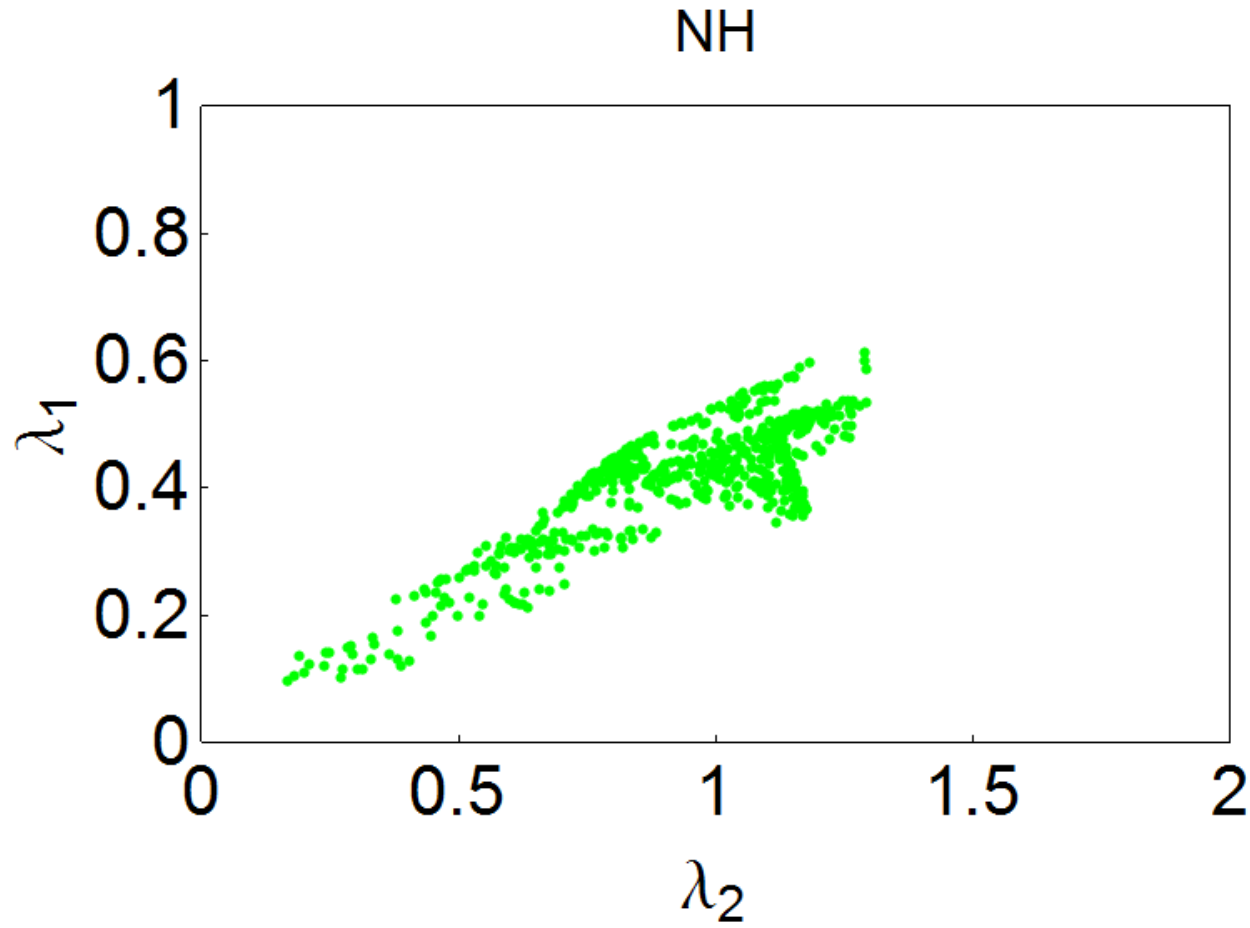}
 \hspace{0.2 truein}
\includegraphics[width=6.0cm,height=5.0cm]{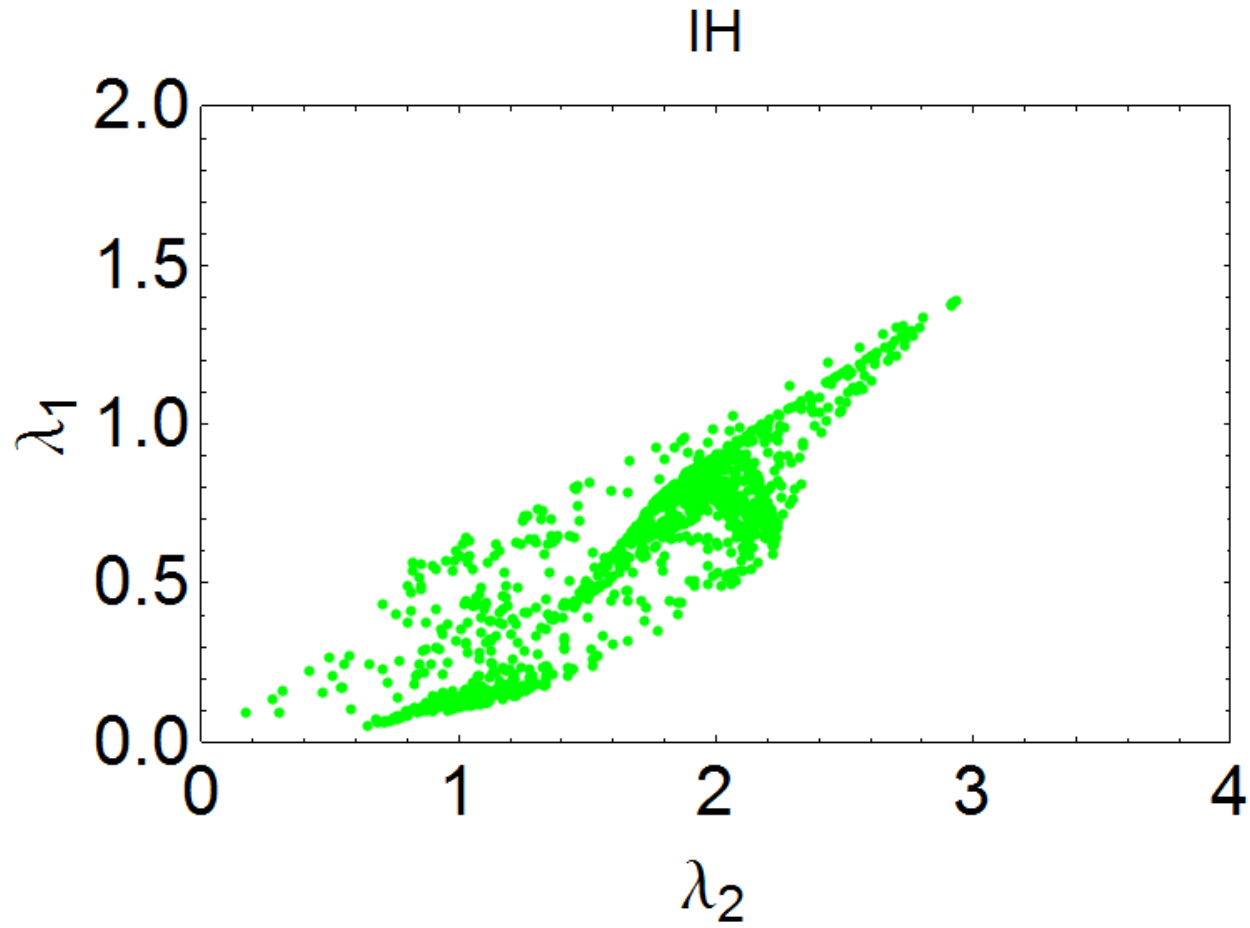}\\
\includegraphics[width=6.0cm,height=5.0cm]{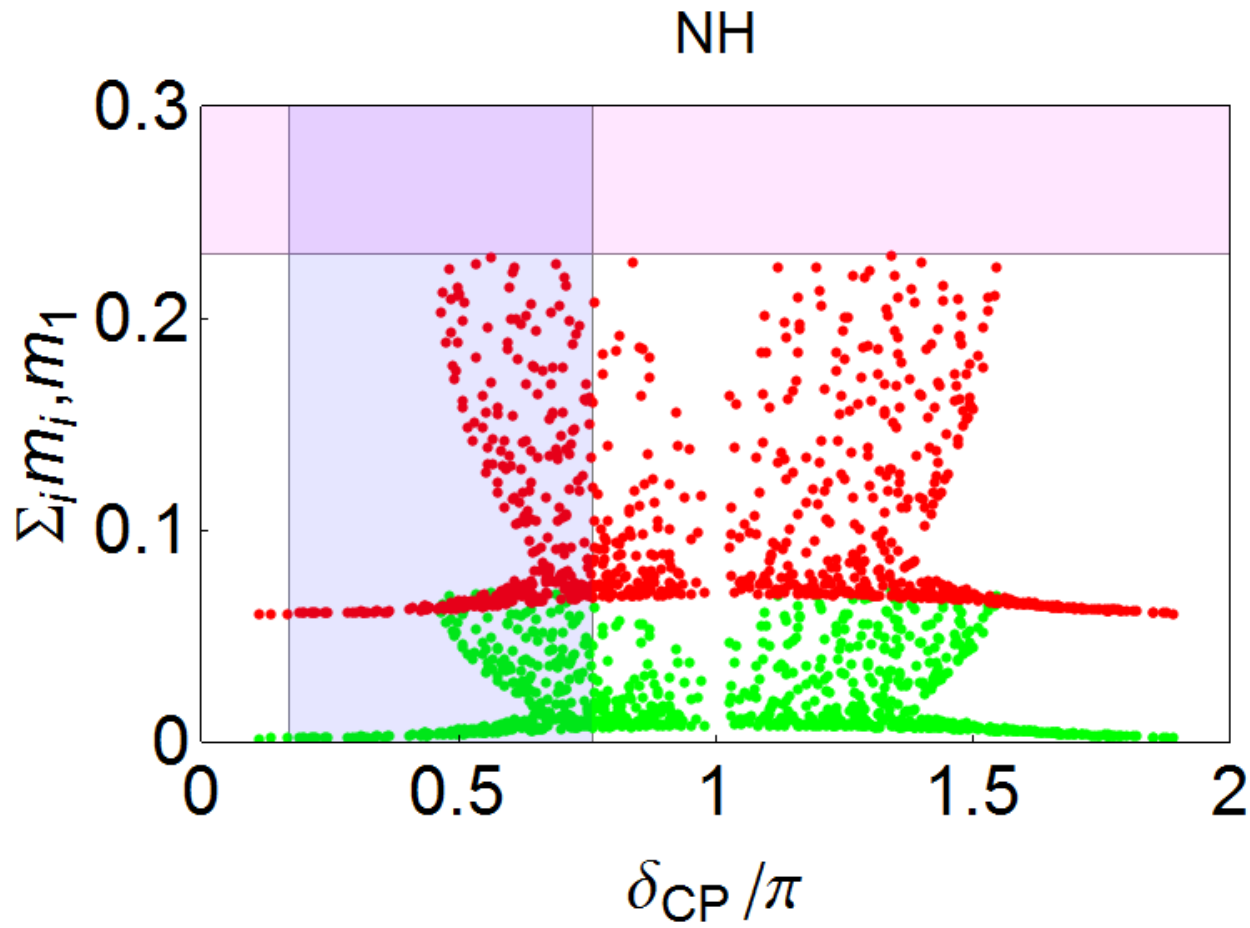}
 \hspace{0.2 truein}
\includegraphics[width=6.0cm,height=5.0cm]{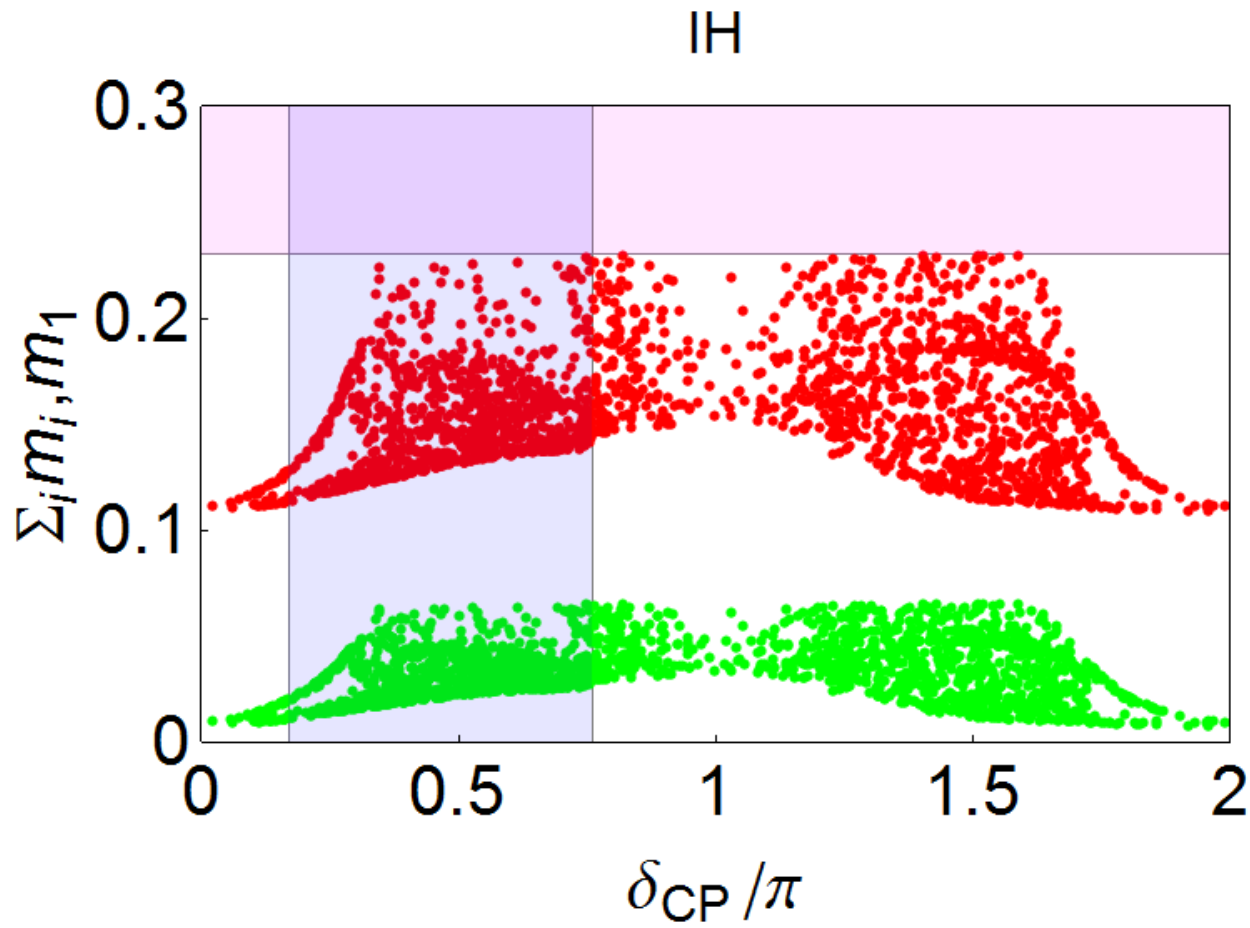}
\caption{Correlation plots between $\lambda_1$ and $\lambda_2$ for normal hierarchy (top left panel), for inverted hierarchy (top right panel) and 
between $\Sigma_im_i$, $m_i$ and $\delta_{CP}$ in the bottom left (right) panel for normal (inverted) hierarchy. The vertical and horizontal bands 
represents the values of $\delta_{CP}$ beyond its $3\sigma$ range and $\Sigma_i m_i>0.23$ eV, the upper bound on sum of active neutrino masses 
given by Planck data, respectively.}\label{c2}
\end{figure}
 \begin{figure}[!htb]
 \includegraphics[width=6.0cm,height=5.0cm]{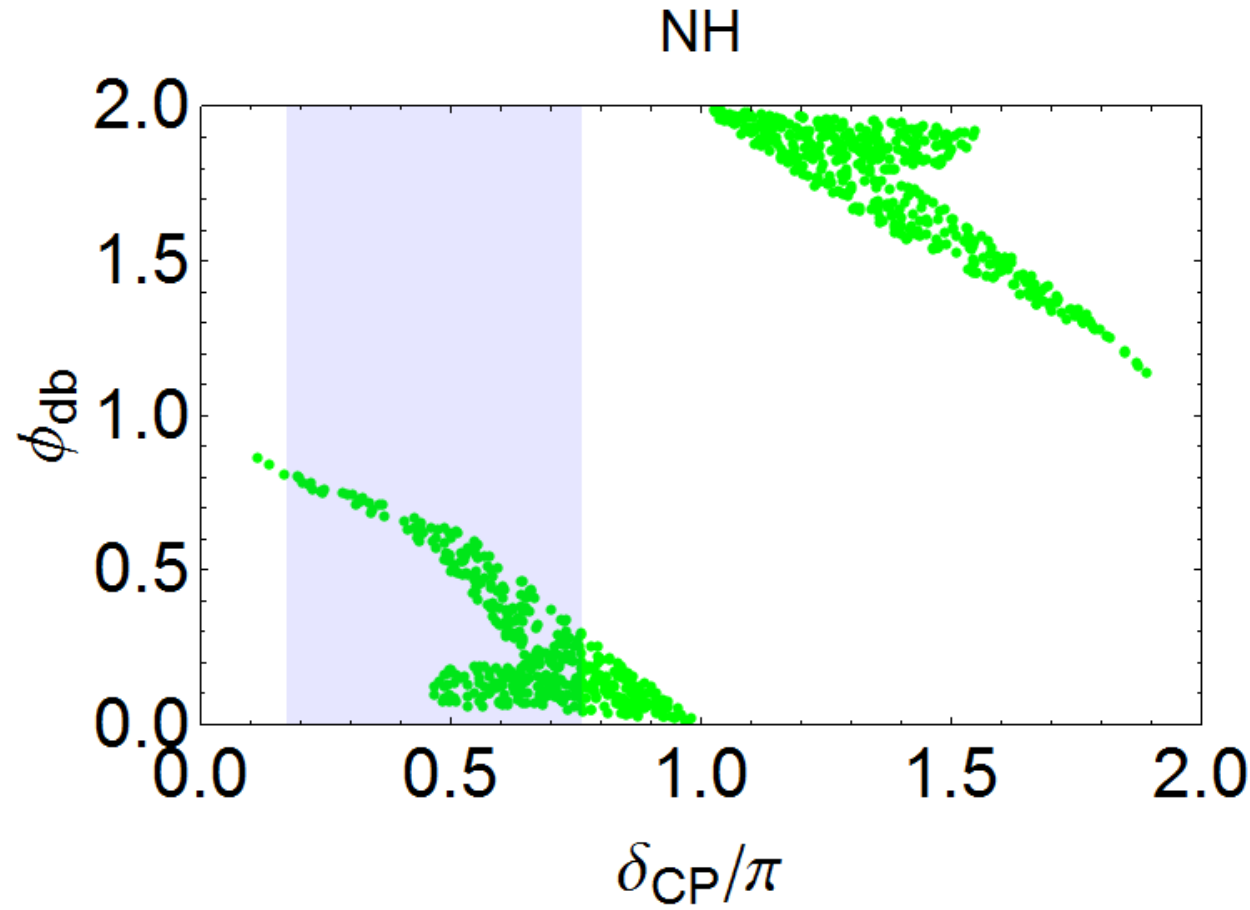}
 \hspace{0.2 truein}
\includegraphics[width=6.0cm,height=5.0cm]{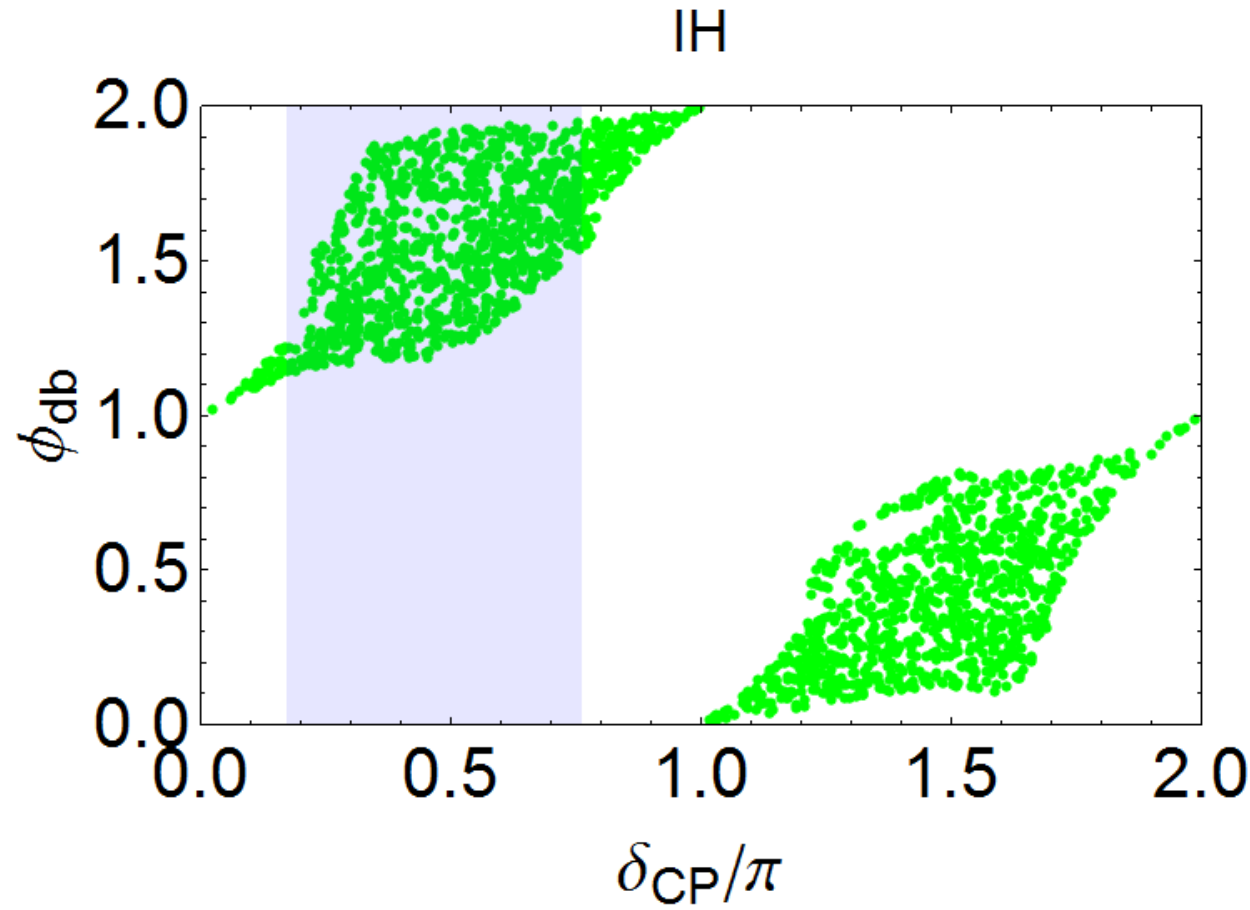}\\
\includegraphics[width=6.0cm,height=5.0cm]{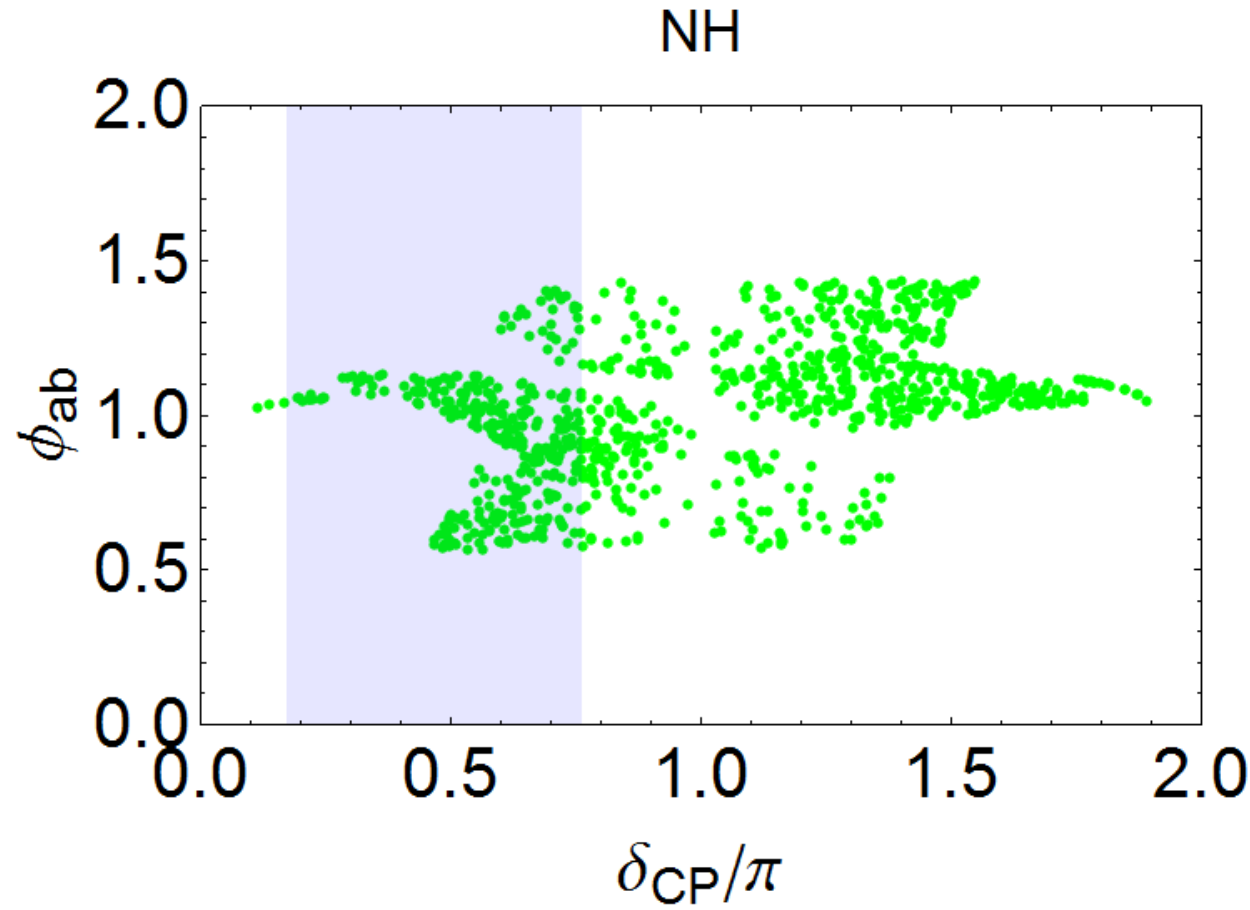}
\hspace{0.2 truein}
\includegraphics[width=6.0cm,height=5.0cm]{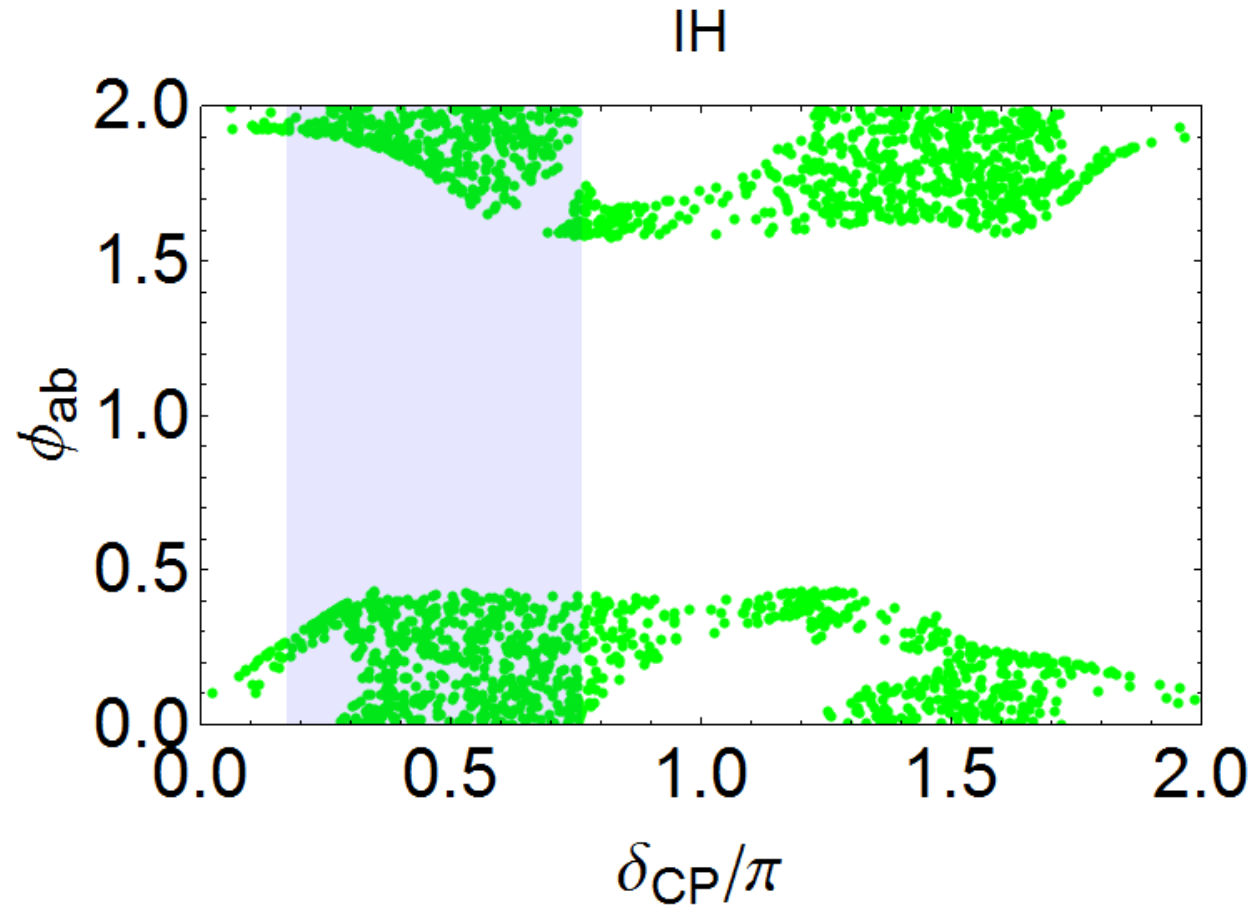}\\
\includegraphics[width=6.0cm,height=5.0cm]{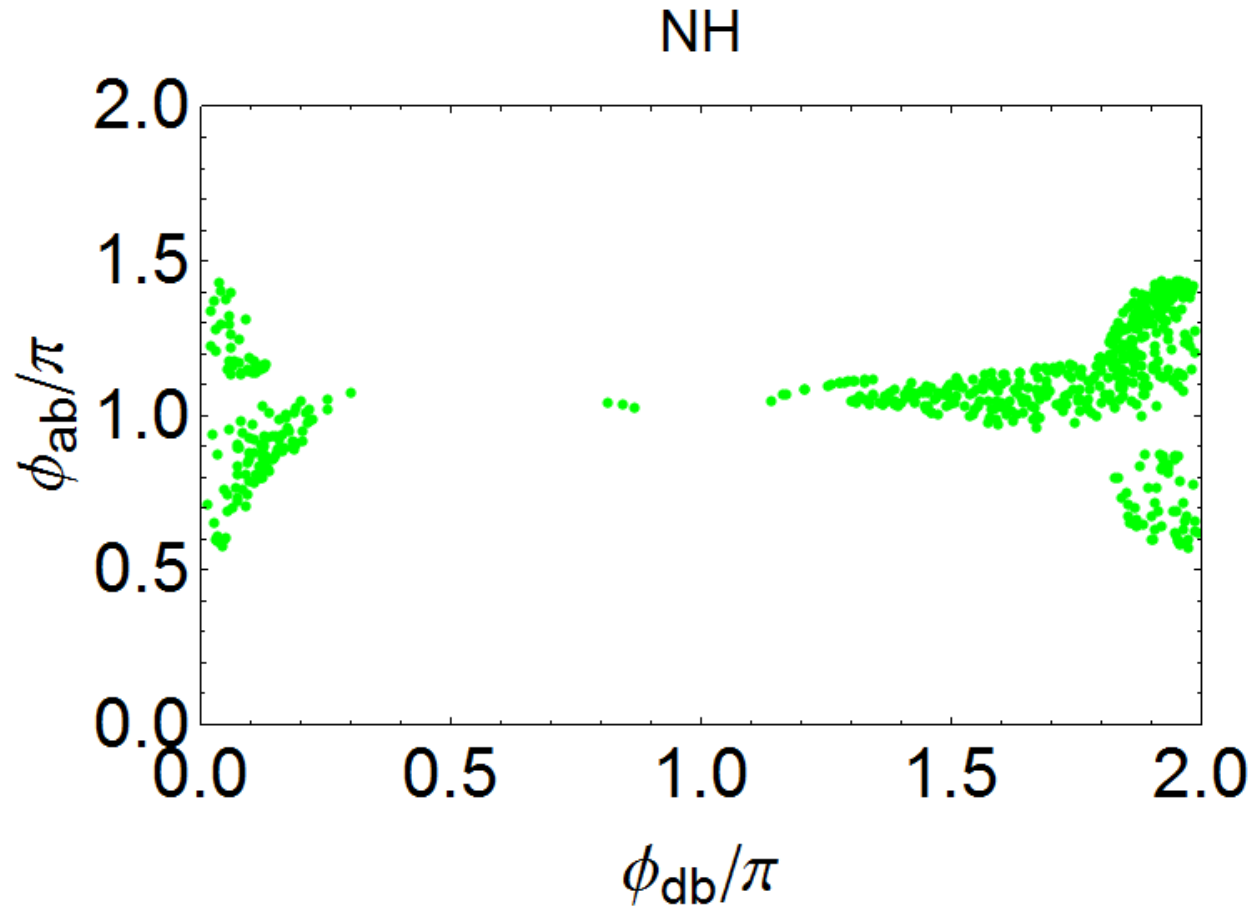}
\hspace{0.2 truein}
\includegraphics[width=6.0cm,height=5.0cm]{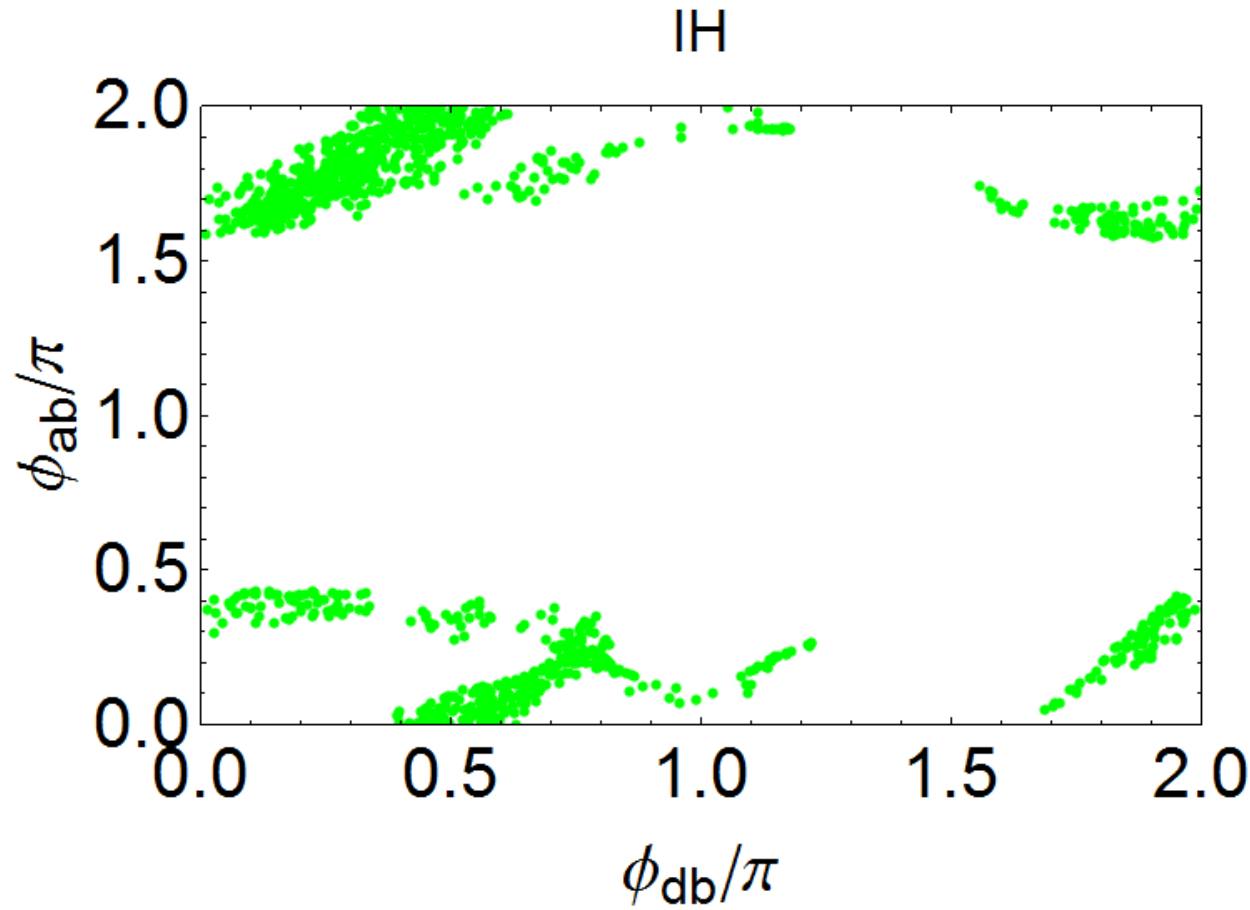}
\caption{Correlation plots between $\phi_{db}$, $\phi_{ab}$ and $\delta_{CP}$ for normal (left panel) and inverted (right panel) hierarchy. The vertical band represents the values of $\delta_{CP}$ beyond its $3\sigma$ range.}\label{c3}
\end{figure}
\subsection{Correlation between model parameters with $\tan\psi \neq 0$.}
With $\tan\psi \neq 0$, the analytic expression for $\lambda_1$ is given by
 \begin{equation}
 \lambda_1=\frac{2\lambda_2\tan 2\theta\cos\phi_{ab}\sin\psi}{\sin\phi_{ab}\left[\sqrt{3}+\tan 2\theta\cos\psi\right]}\;.
 \end{equation}
 We obtain the correlation plots between various parameters as given in Figs. \ref{c2} and Fig. \ref{c3}, by varying $\phi_{db}$ between 
 0 to $2\pi$ and $\delta_{CP}$ in its $3\sigma$ range $(0-0.17\pi \oplus 0.76\pi-2\pi)$ while fixing $\sin^2\theta_{13}$ at its best fit value \cite{Capozzi:2017ipn}.
 
 \newpage

\begin{figure}[!h]
\includegraphics[width=7cm,height=6.0cm]{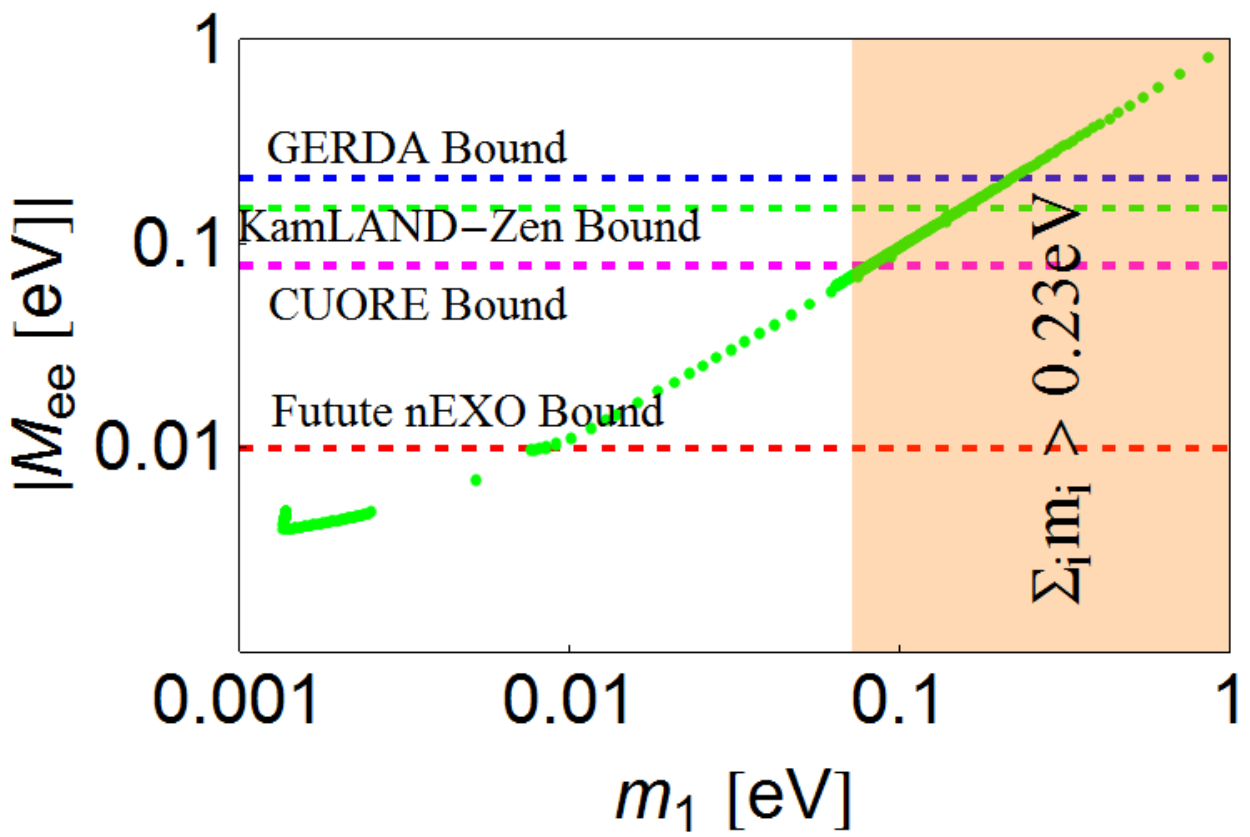}
\hspace{0.2 truein}
\includegraphics[width=7cm,height=6.0cm]{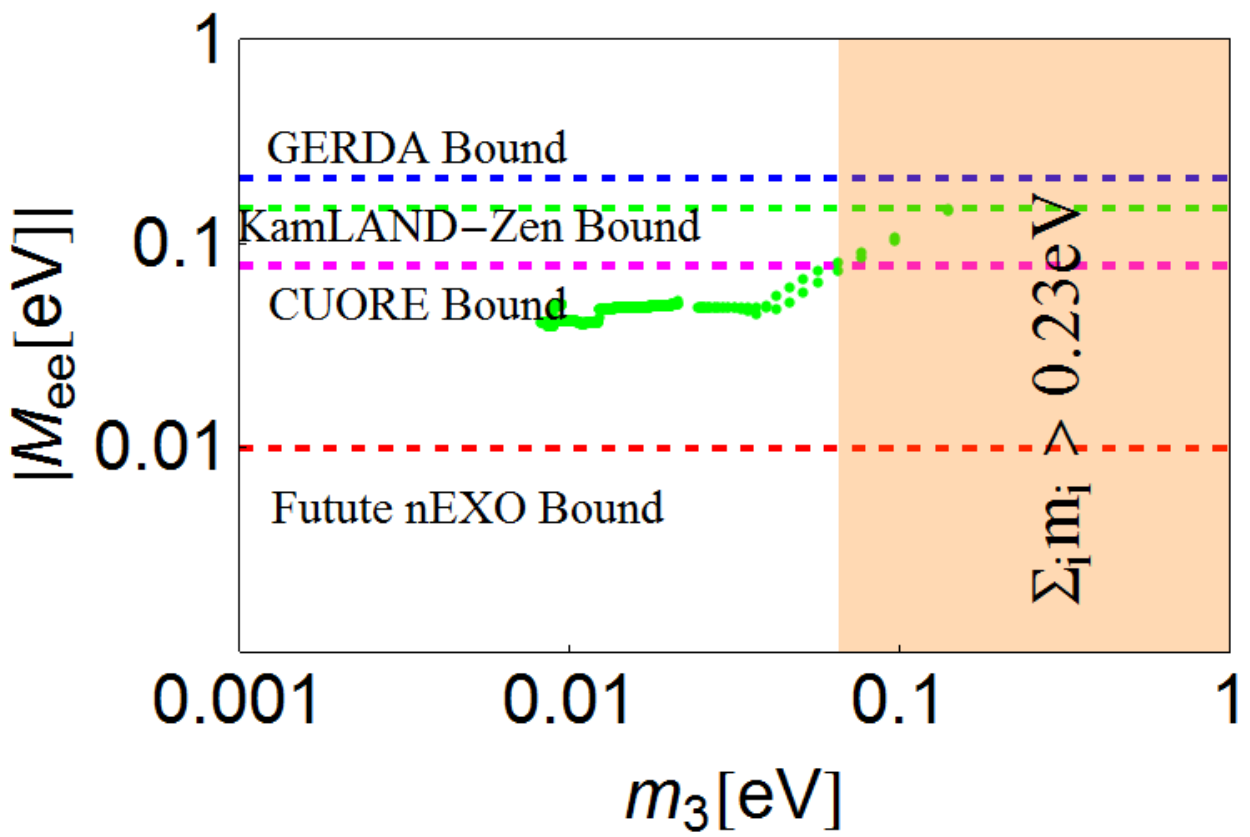}
\caption{Variation of  Majorana parameter $M_{ee}$ which is an observable in neutrino less double beta decay with lightest neutrino mass for the case of normal(left panel) and inverted hierarchy(right panel) for $\tan \psi=0$. }\label{n01}
\end{figure}

{ \it Comment on Neutrinoless double beta decay:}  
 The experimental observation of neutrinoless double beta decay would not only ascertain the Lepton Number Violation (LNV) in nature but it can 
 also give absolute scale of lightest active neutrino mass. The experimental non-observation of such a event puts a bound on half-life of this process on various isotopes 
 which can be translated as a bound on particle physics parameter called as Effective Majorana Mass. In the linear seesaw model, the light Majorana neutrinos 
 contribute to neutrinoless double beta decay while the heavy pseudo-Dirac neutrinos give suppressed contribution.
  
 The measure of LNV can be understood with the key parameter called Effective Majorana Mass which is defined as
\begin{subequations}
\begin{eqnarray}
\left| M_{ee} \right| \equiv \left| m^\nu_{ee} \right|&=&
  \bigg| U^2_{e1}\, m_1 + U^2_{e2}\, m_2 e^{i \rho} + U^2_{e3}\, m_3 e^{i \sigma} \bigg|.  
\label{eq:mee-std}
\end{eqnarray}
\end{subequations}
The light neutrino mass eigenvalues $m_1, m_2, m_3$ depend on input model parameters. 
These input model parameters are constrained to their allowed range in order to satisfy
the oscillation data giving correct values of mass-squared differences and mixings. The 
Majorana phases $\rho$ and $\sigma$ are related to $\phi_{ab}$ and $\phi_{db}$ in some way 
and thus, they are constrained to take limited value. 
The element of PMNS mixing matrix derived from the knowledge of tribimaximal mixing multiplied 
by rotation matrix in $13$ plane. The estimation of Effective Majorana 
mass parameter using these already constrained input model parameters with the variation of 
lightest neutrino mass in displayed in Fig.\ref{n01} where left-panel is for NH and right-panel 
is for IH pattern of light neutrino masses.

\begin{figure}[!htb]
 \includegraphics[width=7.0cm,height=6.0cm]{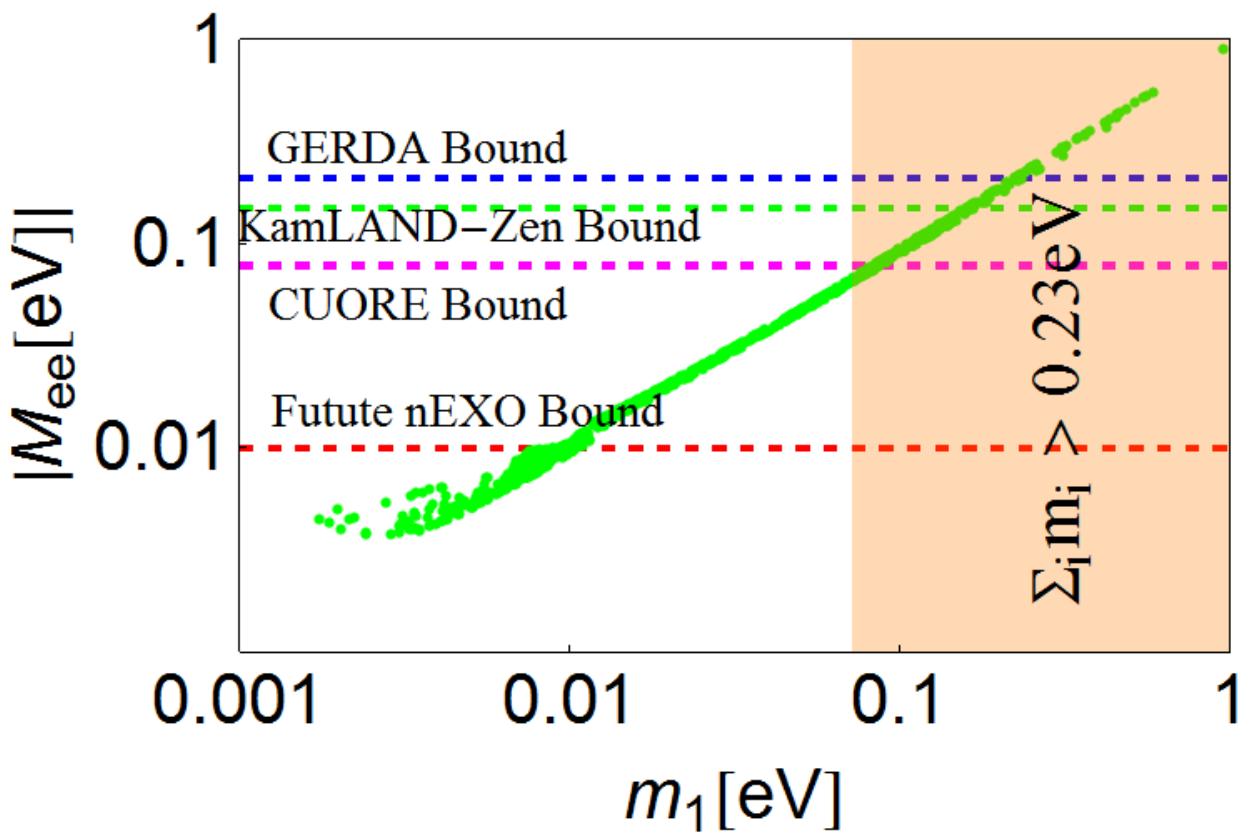}
 \hspace{0.2 truein}
\includegraphics[width=7.0cm,height=6.0cm]{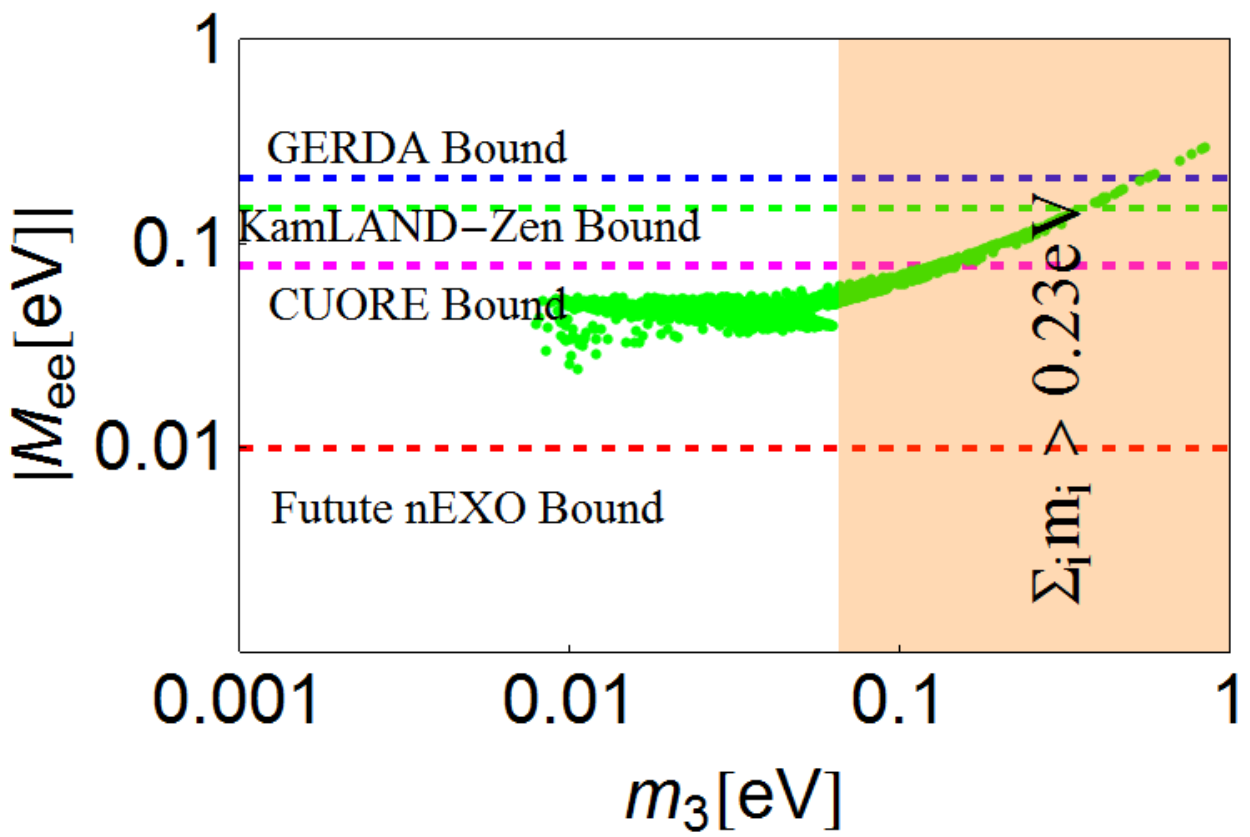}
\caption{Variation of effective Majorana parameter $M_{ee}$ which is a measure of lepton number violation with lightest neutrino mass for the case of normal (left panel) 
and inverted hierarchy (right panel) for $\tan \psi \neq 0$. }\label{0n2}
\end{figure}

The current limit on half-life (or translated bound on Effective Majorana Mass parameter $m^\nu_{ee}$) from GERDA Phase-I~\cite{Agostini:2013mzu} is $T_{1/2}^{0\nu}(^{76}\text{Ge}) > 2.3\times 10^{25}$~yr implies $|m_{ee}| \leq \mbox{0.21\,eV}$ and from KamLAND-Zen~\cite{KamLAND-Zen:2016pfg} as $T_{1/2}^{0\nu}(^{136}\text{Xe}) > 1.07\times 10^{26}$~yr implies $|m_{ee}| \leq \mbox{0.15\,eV}$. There is also bound from CUORE experiment on effective Majorana mass 
parameter as $|m_{ee}| \leq \mbox{0.073\,eV}$~\cite{DellOro:2016tmg}. The expected reach of the future planned $0\nu\beta\beta$ experiments
including nEXO experiment gives $T_{1/2}^{0\nu}(^{136}\text{Xe}) \approx 6.6\times 10^{27}$~yr \cite{Albert:2014afa}.
The variation of Effective mass parameter in green points with lightest neutrino mass is shown in Fig.\ref{n01} 
for $\tan \psi=0$ and the same is plotted in Fig. \ref{0n2} for $\tan \psi \neq 0$. The left-panel is for NH pattern and right-panel is for IH patten of light neutrino masses. The horizontal lines represent the bounds on Effective Majorana mass from various neutrinoless double beta decay experiments while the vertical shaded region are disfavored from Planck data. The present bound is $m_\Sigma < 0.23$~eV from Planck+WP+highL+BAO 
data (Planck1) at 95\% C.L. and $m_\Sigma < 1.08$~eV from Planck+WP+highL (Planck2) at 95\% C.L. \cite{Ade:2013zuv,Ade:2015xua}. 

This plot shows that quasi-degenerate pattern of light neutrinos are disfavoured if we consider the bound on sum of light neutrino masses from cosmology. The current bound on Effective mass parameters from GERDA Phase-I and KamLAND-Zen 
proves that NH and IH pattern of light neutrinos are not sensitive. However, the future planned nEXO experiment is sensitive to both pattern of light neutrinos. 

\section{Leptogenesis}
It is well known that leptogenesis is one of the most elegant frameworks for dynamically generating the observed baryon asymmetry of the Universe. In the resonance leptogenesis scenarios, 
since the mass difference between two or more heavy neutrinos is much smaller than their masses and comparable to their widths, the CP asymmetry in their decays occurs primarily through 
self-energy effects ($\epsilon$-type) rather than vertex effect ($\epsilon'$-type) and gets resonantly enhanced. In the present $A_4$ realization, since the mass splitting between the two heavy 
neutrinos is rather tiny, it provides the opportunity for resonant leptogenesis, which will be discussed in this section.

During the calculation of light neutrino masses and mixing, we have neglected the higher order terms in the Lagrangian ${\cal L}_{\nu}$ as displayed in Eq.(\ref{a4lag}), which are given with 
extra dimension six operators as follows
\begin{equation}
-\left\{\left[\lambda_{N\phi}\phi_S+\lambda_{N\xi}\xi+\lambda_{N\xi^{\prime}}
\xi^{\prime}
\right]\frac{\rho\rho^{\prime}}{\Lambda^2}\overline{N}_RN_R^c+
\left[\lambda_{S\phi}\phi_S^{\dagger}+\lambda_{S\xi}\psi^{\dagger}+
\lambda_{S\xi^{\prime}}{\xi^{\prime}}^{\dagger}\right]\frac{\rho{\rho^{\prime}}^{\dagger}}{\Lambda^2}\overline{S}_R S_R^c
\right\},
\end{equation} 
as these extra terms do not make much difference in those calculations, but they make tiny mass splitting in doubly degenerate mass states of heavy neutrinos. Including these additional terms,  the Majorana mass matrix $\mathbb{M}_2$ becomes
\begin{equation}
\mathbb{M}_2=\left(
\begin{array}{cc}
 m_R &m_{RS} \\
  m_{RS}^T & m_S
\end{array}
\right),
\end{equation}
where 
\begin{eqnarray}\nonumber
m_R&=&\frac{v_{\rho}v_{\rho^{\prime}}}{\Lambda^2}
\left(\begin{array}{c c c}
\frac{2}{3}\lambda_{N\phi}v_S+\lambda_{N\xi}v_{\xi}&-\frac{1}{3}\lambda_{N\phi}v_S&-\frac{1}{3}\lambda_{N\phi}v_S\\
-\frac{1}{3}\lambda_{N\phi}v_S &\frac{2}{3}\lambda_{N\phi}v_S &
-\frac{1}{3}\lambda_{N\phi}v_S+\lambda_{N\xi}v_{\xi} \\
-\frac{1}{3}\lambda_{N\phi}v_S &-\frac{1}{3}\lambda_{N\phi}v_S+\lambda_{N\xi}v_{\xi} &\frac{2}{3}\lambda_{N\phi}v_S
\end{array}
\right)\\ \nonumber
 &+&
{\frac{v_{\rho}v_{\rho^{\prime}}}{\Lambda^2}}
\left(\begin{array}{ccc}
0 & 0 &\lambda_{N\xi^{\prime}}v_{\xi^{\prime}} \\
0&\lambda_{N\xi^{\prime}}v_{\xi^{\prime}}&0\\
\lambda_{N\xi^{\prime}}v_{\xi^{\prime}}&0 &0
\end{array}
\right)\;,
\end{eqnarray}
and
\begin{eqnarray}\nonumber
m_S &=&\frac{v_{\rho}v_{\rho^{\prime}}}{\Lambda^2}
\left(\begin{array}{c c c}
\frac{2}{3}\lambda_{S\phi}v_S+\lambda_{S\xi}v_{\xi}&-\frac{1}{3}\lambda_{S\phi}v_S&-\frac{1}{3}\lambda_{S\phi}v_S\\
-\frac{1}{3}\lambda_{S\phi}v_S &\frac{2}{3}\lambda_{S\phi}v_S &
-\frac{1}{3}\lambda_{S\phi}v_S+\lambda_{S\xi}v_{\xi} \\
-\frac{1}{3}\lambda_{S\phi}v_S &-\frac{1}{3}\lambda_{S\phi}v_S+\lambda_{S\xi}v_{\xi} &\frac{2}{3}\lambda_{S\phi}v_S
\end{array}
\right)\\ 
&+&
{\frac{v_{\rho}v_{\rho^{\prime}}}{\Lambda^2}}
\left(\begin{array}{ccc}
0 & 0 &\lambda_{S\xi^{\prime}}v_{\xi^{\prime}} \\
0&\lambda_{S\xi^{\prime}}v_{\xi^{\prime}}&0\\
\lambda_{S\xi^{\prime}}v_{\xi^{\prime}}&0 &0
\end{array}
\right)\;.
\end{eqnarray}
The mass matrix $\mathbb{M}_2$ can be  approximately block diagonalized by the unitary matrix $\frac{1}{\sqrt{2}} \left(
\begin{array}{cc}
 I &-I \\
 I &I
\end{array}
\right)$ and becomes
\begin{eqnarray}
\mathbb{M}_2^{\prime}=\left(
\begin{array}{cc}
 m_{RS}+\displaystyle{\frac{m_R+m_S}{2}} &m_S-m_R \\
 m_S-m_R &-m_{RS}+\displaystyle{\frac{m_R+m_S}{2}}
\end{array}
\right)
\approx  \left(
\begin{array}{cc}
 m_{RS}+\displaystyle{\frac{m_R+m_S}{2}} &0 \\
 0 &-m_{RS}+\displaystyle{\frac{m_R+m_S}{2}}
\end{array}
\right),\hspace{0.5 truecm}
\end{eqnarray}
with eigenvalues 
\begin{eqnarray}\nonumber
{M_1^{\prime}}^{\pm}\approx M_1\left(1\pm\frac{v_{\rho}v_{\rho^{\prime}}}{\Lambda^2}\frac{m_1^{\prime}}{M_1}\right)\;, \nonumber\\
{M_2^{\prime}}^{\pm}\approx M_2\left(1\pm\frac{v_{\rho}v_{\rho^{\prime}}}{\Lambda^2}\frac{m_2^{\prime}}{M_2}\right)\;, \nonumber\\
{M_3^{\prime}}^{\pm}\approx M_3\left(1\pm\frac{v_{\rho}v_{\rho^{\prime}}}{\Lambda^2}\frac{m_3^{\prime}}{M_3}\right)\;,
\end{eqnarray}
where
\begin{eqnarray}
m_1^{\prime}&=& 2\text{Re}\left\lbrace\left[a^{\prime}-\left(\frac{bb^{\prime}-\frac{1}{2}\left(bd^{\prime}+b^{\prime}d\right)+dd^{\prime}}{\sqrt{b^2-bd+d^2}}\right)\right]e^{-i\phi_1}\right\rbrace\;, \nonumber \\
m_2^{\prime}&=&2\text{Re}\left[\left(b^{\prime}+d^{\prime} \right)e^{-i\phi_2}\right]\;,\nonumber\\
m_3^{\prime}&=&2\text{Re}\left\lbrace\left[a^{\prime}+\left(\frac{bb^{\prime}-\frac{1}{2}\left(bd^{\prime}+b^{\prime}d\right)+dd^{\prime}}{\sqrt{b^2-bd+d^2}}\right)\right]e^{-i\phi_3}\right\rbrace\;, \nonumber \\
a^{\prime}&=&\frac{1}{2}\left(\lambda_{N\phi}+\lambda_{S\phi}\right)v_S ,~~
b^{\prime}=\frac{1}{2}\left(\lambda_{N\xi}+\lambda_{S\xi}\right)v_{\xi}
\;,\nonumber \\
d^{\prime}&=&\frac{1}{2}\left(\lambda_{N\xi^{\prime}}
+\lambda_{S\xi^{\prime}}\right)v_{\xi^{\prime}}\;.
\end{eqnarray}
 and $\phi_i$ is the phase associated with $\tilde{M}_i$.   The above set of  equations show that $m_i^{\prime}$ can be of the order of $M_i$ since $a$, $a^{\prime}$ are of the order of $v_S$, $b$, $b^{\prime}$ are of the order of $v_{\xi}$ and $d$, $d^{\prime}$ are of the order of $v_{\xi^{\prime}}$.
 
The  decay of nearly degenerate heavy neutrinos creates lepton asymmetry, and is given as  \cite{Gu:2010xc}
\begin{eqnarray}
\epsilon_{N_i^{\pm}}=&-&\frac{1}{4\pi A_{N_i^{\pm}}} \left[\displaystyle{\left(\frac{\tilde{m}_{D}}{v}\right)^{\dagger}\left(\frac{\tilde{m}_{D}}{v}\right)-
\left(\frac{\tilde{m}_{LS}}{v}\right)^{\dagger}\left(\frac{\tilde{m}_{LS}}{v}\right)}
\right]_{ii} \displaystyle{ \text{Im}\left[\frac{\tilde{m}_{D}^{\dagger}\tilde{m}_{LS}}{v^2}\right]}_{ii} \nonumber\\
&\times &  \frac{r_{N_i}}{\displaystyle{{r_{N_i}}^2+
\frac{1}{64\pi^2}{A_{N_i^{\pm}}}^2}}\;,
\end{eqnarray}
where
\begin{eqnarray}
A_{N_i^{\pm}}&=&\frac{1}{2}\left[\left(\frac{{\tilde{m}_D}^{\dagger}}{v}\pm \frac{{\tilde{m}_{LS}}^{\dagger}}{v}\right)
\left(\frac{{\tilde{m}_D}}{v}\pm \frac{{\tilde{m}_{LS}}}{v}\right)\right]_{ii}\\ \nonumber
r_{N_i}&=&\frac{{{M_i^{\prime}}^+}^2-{{M_i^{\prime}}^-}^2}{{{M_i^{\prime}}^+}{{M_i^{\prime}}^-}}\approx 4\left(\frac{v_{\rho}v_{\rho^{\prime}}}{\Lambda^2}\frac{m_i^{\prime}}{M_i}\right)\;,\nonumber \\
\nonumber\\
\tilde{m}_D&=&m_D U_\text{TBM} U_{13}^T,~~~~\tilde{m}_{LS}=m_{LS}U_\text{TBM}U_{13}^T\;.
\end{eqnarray}
Since $r_{N_i}\ll A_{N_i^{\pm}}$, ${r_{N_i}}^2+\frac{1}{64\pi^2}A_{N_i^{\pm}}^2\approx \frac{1}{64\pi^2}A_{N_i^{\pm}}^2$, for $\tilde{m}_{LS}\ll\tilde{m}_D$ 
\begin{equation}
\epsilon_{N_i^{\pm}}\approx -128\pi \text{Im}\left[\tilde{m}_{LS}^{\dagger}\tilde{m}_D\right]_{ii}\frac{r_{N_i}v^2}{\left(\tilde{m}_D^{\dagger}\tilde{m}_D\right)^2}\;.
\end{equation}
Substituting  $\tilde{m}_D^{\dagger}\tilde{m}_D=|y_1|^2\left (v v_{\rho^{\prime}}/\Lambda \right)^2$, $\tilde{m}_{D}^{\dagger}\tilde{m}_{LS}=y_1^*y_2 v^2 \left (v_{\rho} v_{\rho^{\prime}}/{\Lambda}^2 \right )$ and  $r_{N_i}\approx 4\left(v_{\rho}v_{\rho^{\prime}}/{\Lambda^2}\right ) \left (m_i^{\prime}/{M_i}\right)$ in the above equation, we obtain
\begin{equation}
\epsilon_{N_i^{\pm}}\approx -512\pi\left(\frac{v_{\rho}}{v_{\rho^{\prime}}}\right)^2 \frac{\text{Im}\left[y_1^*y_2 \right ]}{|y_1|^4}\frac{m_i^{\prime}}{M_i}\;.
\end{equation}
Writing $y_1^*y_2=|y_1y_2|e^{i\theta_{\epsilon}}$, one can have
\begin{equation}
\epsilon_{N_i^{\pm}}\approx -512\pi\left(\frac{v_{\rho}}{v_{\rho^{\prime}}}\right)^2 \frac{|y_2|}{|y_1|^3}\frac{m_i^{\prime}}{M_i}\sin \theta_{\epsilon}\;.
\end{equation}

Here we  calculate the  baryon asymmetry for the case $M_3 \ll M_2< M_1$, i.e., normal hierarchy in active neutrino sector.  
It is mainly the decay of $M_3^{\pm}$ that contributes to the final baryon asymmetry. Since the decay is in strong wash out region, 
final baryon asymmetry is given by \cite{Gu:2010xc},
\begin{equation}
\eta_B=-\frac{28}{79}\left(\frac{0.3\epsilon_{N_3^{\pm}}}{g_*K_{N_3^{\pm}}\left(\ln K_{N_3^{\pm}}\right)^{0.6}}\right),\label{cp}
\end{equation}
  where $K_{N_i^{\pm}}=\displaystyle{\frac{1}{8\pi} \left(\frac{8\pi^3 g_*}{90}\right)^{-1/2}  }\left( \frac{M_{Pl}}{ M_{N_i^{\pm}}}\right )A_{N_i^{\pm}}$,    
   $g_*\approx 106.75$ and $M_\text{Pl}=2.435\times 10^{18}~\text{GeV}$ are relativistic degrees of freedom of SM particles and Planck mass respectively. Here 
\begin{equation}
K_{N_3^{\pm}}=K_{N_3}\approx 0.234\left[\frac{m_3~(\text{(eV)}}{10^{-2}}\right]\frac{v_{\rho^{\prime}}}{v_{\rho}} \gg 1 \;,\label{b1}
\end{equation}  
 as $m_3$ is of the order of $10^{-2} ~\text{eV}$ and $\frac{v_{\rho^{\prime}}}{v_{\rho}}\gg1$. Substituting $K_{N_3^{\pm}}$ and $\epsilon_{N_3^{\pm}}$ in Eqn. (\ref{cp}) gives
\begin{equation}
\eta_B\approx 0.174\left(\frac{\left(\frac{m_3(\text{eV})}{10^{-2}}\right)^2}{K_{N_3}^3(\ln K_{N_3})^{0.6}}\right)\frac{|y_2|m_3^{\prime}}{|y_1|^3M_3}\sin\theta_{\epsilon}\;.
\end{equation}
For $y_1\approx y_2$ and $\frac{m_3^{\prime}}{M_3}\approx 1$ the above equation gives 
\begin{equation}
\eta_B\leq 0.174\left(\frac{\left(\frac{m_3(\text{eV})}{10^{-2}}\right)^2}
{|y_1|^2K_{N_3}^3(\ln K_{N_3})^{0.6}}\right).\label{b2}
\end{equation}

For $m_1<0.005~{\rm eV}$, $m_3\approx 0.05~\text{eV}$, with this value of $m_3$ ,$|y_1|^2=10^{-3}$ and $\eta_B=6.9\times 10^{-10}$ from
 \ref{b1} and \ref{b2} we found the minimum value of $\frac{v_{\rho}}{v_{\rho^{\prime}}}$ requires to generate observed baryon asymmetry as
 \begin{equation}
 \left.\frac{v_{\rho}}{v_{\rho^{\prime}}}\right|_{min}=5.07\times 10^{-5}\;.\label{r1}
 \end{equation}

\noindent
{\it \bf Comment on Non-unitarity in leptonic sector:}\\
In usual case, the light active Majorana neutrino mass matrix is diagonalized by the PMNS mixing 
matrix $U_{\rm PMNS}$ as $U_{\rm PMNS}^{\dagger}\, m_{\nu}\, U^*_{\rm PMNS} = \text{diag}
\left(m_{1},m_{2},m_{3}\right)$ where $m_1,m_2,m_3$ are mass eigenvalues for light neutrinos. 
However, the diagonalizing mixing matrix in case of linear seesaw mechanism--where the neutral 
lepton sector is comprising of light active Majorana neutrinos plus additional two types of 
right-handed sterile neutrinos--is given by
\begin{equation} 
{\cal N}\simeq(1-\eta)U_{\rm PMNS}\, , 
\end{equation}
where the non-unitarity effect is parametrized as \cite{Forero:2011pc},
\begin{equation}
\eta=\frac{1}{2}m_D^*{m_{RS}^{\dagger~ -1}} m_{RS}^{-1}m_D^T\;.\label{et}
\end{equation} 
In the linear seesaw framework under consideration, the $N-S$ mixing matrix $m_{RS}$ is symmetric and with $y_1\approx y_2$,  the 
$\nu-S$ mass term can be expressed as $m_{LS}^{\dagger}m_{LS}=\frac{1}{2}m_lM_0 (v_{\rho}/v_{\rho^{\prime}})$ where $m_0$ and 
$M_0$ are the masses of heaviest active and lightest heavy neutrinos respectively.  Thus, the above relation for $\eta$ can be written in terms 
of light neutrino mass matrix and other input model parameters as
 \begin{equation}
 \eta=\frac{m_{\nu}^*m_{\nu}^T}{4m_0M_0\frac{v_{\rho}}{v_{\rho{\prime}}}}\;.
 \end{equation}
 
The maximum value of $\eta$  for inverted mass hierarchy with lightest neutrino mass $m_l \simeq 0.005~{\rm eV}$ while considering the constrained 
value of the ratio of VEV $\frac{v_{\rho}}{v_{\rho^{\prime}}}=5.07\times 10^{-5}$ as derived from the discussion of leptogenesis and  using $M_0=5~\text{TeV}$ can be obtained as follows 
 \begin{equation}
|\eta|\approx \frac{1}{2}\left[\begin{array}{c c c}
4\times 10^{-12} & 10^{-11} & 10^{-11}\\
 10^{-11} & 5\times 10^{-11} &5\times 10^{-11}\\
10^{-11} & 5\times 10^{-11} &5\times 10^{-11}
 \end{array}\right].
 \end{equation}
Using the representative set of model parameters $m_0$ and $M_0$, the mass matrices $m_D$ and $m_{LS}$ are expressed as follows
 \begin{equation}
 m_D=\sqrt{\frac{M_0m_0}{v_{\rho}/v_{\rho^{\prime}}}},~~~ m_{LS}=\sqrt{\frac{v_{\rho}}{v_{\rho^{\prime}}}M_0m_0}\;.
 \end{equation}
 Using the constrained value of these model parameters $m_0$ and $M_0$, the Dirac neutrino mass connecting $\nu-N$ is found to be $m_D\approx 70 $ MeV and the 
 other mass term connecting $\nu-S$ is $m_{LS}\approx 3.5$ keV.
 \section{Conclusion}
In this paper we have considered the realization of linear seesaw by extending SM symmetry with  $A_4\times Z_4\times Z_2$ along with a global symmetry $U(1)_X$ which is broken 
explicitly in Higgs potential. In addition to SM fermions, the model has six heavy fermions, three right-handed neutrinos $(N_i)$ and three sterile neutrinos $(S_i)$. We found that each 
mass state of heavy neutrino is nearly doubly degenerate with a small mass splitting, which can be neglected for the   calculation of active neutrino mass and mixing parameters. 
The mass of active neutrinos are found to be  inversely proportional to that of heavy neutrinos. The model predicts lepton mixing matrix i.e., the PMNS as $U_{TBM}\cdot U_{13}\cdot P$,  
where $U_{13}$ is the rotation in 13 plane and hence,  explains well the results on mixing angles and $\delta_{CP}$ from oscillation experiments. We obtained the parametric space 
and correlation plots between various observables by fixing $\theta_{13}$ at its best-fit value and the ratio mass squared differences, $\Delta m^2_{21}/\left|\Delta m^2_{13}\right|$ 
at 0.03 and varying $\delta_{CP}$ in its $3\sigma$ range.

We have demonstrated that pairs of nearly degenerate Majorana neutrinos in the  model opens up  the scope to resonant leptogenesis 
to account for the baryon asymmetry of the universe. We calculated the minimum value of $v_{\rho}/v_{\rho^{\prime}}$
to generate observed baryon asymmetry by fixing the mass of lightest heavy neutrino in TeV for the case where heavy neutrino mass are highly hierarchical so that the only contribution 
to baryon asymmetry is from the decay of two lightest heavy neutrinos, the parameter space which satisfies this condition predicts normal hierarchy in active neutrino sector with lightest 
on less than $0.005$ eV. In this case the maximum non-unitarity value, the model can accommodate in leptonic sector is very small and is of the order of $10^{-11}$ and  the mass parameters are found to be $m_D\approx 70 $ MeV 
and $m_{LS}\approx$ 3.5 keV.
 
\acknowledgements
SM  would like to thank University Grants Commission for financial support.  RM acknowledges the  support from the  Science and Engineering Research Board (SERB),
Government of India through grant No. SB/S2/HEP-017/2013. 

\bibliographystyle{utcaps_mod.bst}
\bibliography{lsref.bib}

\end{document}